\documentclass[transmag]{IEEEtran}

\usepackage{lipsum}
\usepackage{amsfonts}
\usepackage{graphicx}
\usepackage{epstopdf}
\usepackage{pgfplots}

\usepackage{multicol}
\usepackage{multirow}
\usepackage{colortbl}
\usepackage{bm}

\usepackage{url}
\usepackage{cite}
\usepackage[hidelinks]{hyperref}
\hypersetup{
    colorlinks=true,
    linkcolor=black,
    filecolor=black, 
    citecolor=black,
    urlcolor=black,
}

\ifCLASSOPTIONcompsoc
    \usepackage[caption=false, font=normalsize, labelfont=sf, textfont=sf]{subfig}
\else
\usepackage[caption=false, font=footnotesize]{subfig}
\fi
\usepackage[export]{adjustbox}

\ifpdf
  \DeclareGraphicsExtensions{.eps,.pdf,.png,.jpg}
\else
  \DeclareGraphicsExtensions{.eps}
\fi

\usepackage{amsopn}
\usepackage{mathtools}

\usepackage[algo2e,ruled,vlined,linesnumbered]{algorithm2e}
\SetInd{0em}{0.5em}
\SetKwComment{Comment}{/\!\!/}{}

\usepgflibrary{shapes.geometric}
\usetikzlibrary{calc}
\pgfplotsset{compat=1.16}

\def\Fig{Fig.~}

\newcommand{\soa}{state-of-the-art }

\newcommand{\R}{\mathbb R}  
\newcommand{\N}{\mathcal N} 

\DeclarePairedDelimiterX{\norm}[1]{\lVert}{\rVert}{#1} 
\DeclarePairedDelimiterX{\abs}[1]{\lvert}{\rvert}{#1}
\DeclarePairedDelimiterX{\ip}[2]{\langle}{\rangle}{#1, #2}
\DeclarePairedDelimiterX{\pfrac}[2]{(}{)}{\frac{#1}{#2}}
\DeclareMathOperator{\sign}{sign}

\DeclareMathOperator{\ST}{\mathrm{ST}} 
\DeclareMathOperator{\divrg}{div}

\let\D\undefined
\newcommand{\D}{\bm{D}}
\let\x\undefined
\newcommand{\x}{\bm{x}}
\let\y\undefined
\newcommand{\y}{\bm{y}}
\let\z\undefined
\newcommand{\z}{\bm{z}}

\let\bnu\undefined
\newcommand{\bnu}{\bm{\nu}}

\let\m\undefined
\newcommand{\m}{\bm{m}}

\let\W\undefined
\newcommand{\W}{\bm{W}}

\begin{document}

\title{CDLNet: Noise-Adaptive Convolutional Dictionary Learning Network for Blind Denoising and Demosaicing}

\author{Nikola Janju\v{s}evi\'{c},~\IEEEmembership{Student Member,~IEEE,}
        Amirhossein~Khalilian-Gourtani,~\IEEEmembership{Student Member,~IEEE,}
        \\and~Yao~Wang,~\IEEEmembership{Fellow,~IEEE}
\thanks{N. Janju\v{s}evi\'{c} is with New York University Tandon School of Engineering, Electrical and Computer Engineering Department, Brooklyn, NY 11201, USA (e-mail: npj226@nyu.edu).}
\thanks{A. Khalilian-Gourtani is with New York University Tandon School of Engineering, Electrical and Computer Engineering Department, Brooklyn, NY 11201, USA (e-mail: akg404@nyu.edu).}
\thanks{Y. Wang is with New York University Tandon School of Engineering, Electrical and Computer Engineering Department, Brooklyn, NY 11201, USA (e-mail: yaowang@nyu.edu).}}

\IEEEtitleabstractindextext{\begin{abstract} Deep learning based methods hold \soa results in low-level image processing tasks, but remain difficult to interpret due to their black-box construction. Unrolled optimization networks present an interpretable alternative to constructing deep neural networks by deriving their architecture from classical iterative optimization methods without use of tricks from the standard deep learning tool-box. So far, such methods have demonstrated performance close to that of \soa models while using their interpretable construction to achieve a comparably low learned parameter count. In this work, we propose an unrolled convolutional dictionary learning network (CDLNet) and demonstrate its competitive denoising and joint denoising and demosaicing (JDD) performance both in low and high parameter count regimes. Specifically, we show that the proposed model outperforms \soa fully convolutional denoising and JDD models when scaled to a similar parameter count. In addition, we leverage the model's interpretable construction to propose a noise-adaptive parameterization of thresholds in the network that enables \soa blind denoising performance, and near-perfect generalization on noise-levels unseen during training. Furthermore, we show that such performance extends to the JDD task and unsupervised learning. \end{abstract}
}

\maketitle

\section{Introduction}

\IEEEPARstart{I}{n}
recent years, deep-learning based methods have become the \soa on
low-level image processing tasks such as denoising, demosaicing, deblurring,
super-resolution, and more. Despite the extensive literature on the theoretical
modeling of the aforementioned tasks, these deep-neural-networks (DNNs) are commonly
constructed as black-box function approximators whose architectures are largely derived
from trial and error \cite{buhrmester2019analysis}. This is made apparent by two wide-spread techniques
for dealing with varying input noise-levels in these imaging tasks: by
simply presenting images of various noise-levels during training \cite{DnCNN} or
by additionally presenting the estimated noise-level as an input to the network
\cite{FFDNet}. In both cases, it is hoped that the network learns a mapping
that is noise-level adaptive.  

A more principled approach to DNN design will lead to better analysis and possible improvement of the network. 
Recent works in the area of principled DNN construction have come in
the form of so-called ``unrolled networks" \cite{unrolling, KoEf20}. These architectures are derived from
iterative optimization algorithms that are truncated and parameterized.
However, many authors parameterize these iterates with black-box neural networks
themselves. While this allows for such interpretations as ``learned
proximal-operators" \cite{ongie2020deep}, it does not offer much insight into DNN construction. 

In this work, we derive a denoising DNN architecture via a \textit{direct parameterization}
of a truncated differentiable convolutional dictionary-learning algorithm, which
we call CDLNet. In doing so, we maintain the defining DNN form of layered linear operations, point-wise non-linearities, and residual connections, while side-stepping the use of batch-normalization, residual learning, and feature-domain processing. From this
interpretable network construction, we are able to naturally account for a
continuous range of input noise-levels by parameterizing the point-wise non-linearities to be a function of the estimated noise-level. We verify the efficacy of our noise-adaptive model by demonstrating
near-perfect generalization of a single model above and below its training
noise-level range. The proposed method may be easily adapted to other imaging
problems with additive Gaussian noise, and a known observation operator, by incorporating said operator into the
network architecture. To this end, we train a single network for joint
denoising and demosaicing (JDD) of natural images which inherits the same
generalization capabilities as the pure denoising network, and performs
competitively to other JDD networks.
We also demonstrate the aforementioned generalization capabilities in the case of 
unsupervised training via a Monte-Carlo Stein's unbiased risk estimator (MC-SURE) loss function.

Previous works have considered DNN architectures similar to CDLNet, however,
they required additional sophisticated modeling -- and ultimately a departure from the
convolutional model -- to achieve competitive performance with \soa DNNs
\cite{Simon2019,Lecouat2020Games,Scetbon2021}. We instead show that local convolutions alone
are sufficient. By increasing the width and depth of our network, along with
efficient management of computation via small-strided convolutions, we are able to
surpass the performance of popular denoising networks like DnCNN \cite{DnCNN}.

Other works have also considered accounting for varying input noise-levels via
augmenting the network parameters \cite{Isogawa2017,Ramzi2020}.
These methods are not derived via interpretable model construction, nor do they demonstrate generalization outside the training noise-level range.

In summary, our contributions are as follows:
\begin{itemize}
	\item We propose CDLNet, a denoising DNN derived from convolutional dictionary learning which employs strided convolutions to enable competitive performance to \soa DNN models with similar complexity (only local operations). 
	\item We propose a noise-adaptive parameterization of the learned thresholds that allows for processing of images from a continuous noise-level input range and generalization to noise-levels unseen during training.
	\item We demonstrate that our method and its generalization capabilities translate well to both the unsupervised learning regime and the task of joint denoising and demosaicing.
\end{itemize}

\subsection{Organization of the Paper}
The rest of our paper is organized as follows. In Section \ref{sec:related}, we review the classical dictionary learning formulation, related works pertaining to unrolled DNNs, and related works on noise-adaptive networks. In Section \ref{sec:ProposedMethod}, we detail our proposed denoising network with noise-level adaptivity. 
In Section \ref{sec:results}, we present experimental results of the proposed networks. 
In Sections \ref{sec:discussion}, \ref{sec:conclusion} we discuss the results and conclude the paper. 

\section{Preliminaries and Related Work} \label{sec:related}

\subsection{Sparse Representation and Dictionary Learning}
Sparse representation has been a driving force in signal processing
and continues to shape methods today. In this model, given a properly designed dictionary (matrix) $\D$, we assume that
a signal of interest $\x \in \R^N$ can be
efficiently composed by only a few columns of $\D$. In the case
of observations contaminated with additive Gaussian white noise
(AWGN), $\y = \x + \bnu$, where $\bnu \sim \N(0,\sigma \boldsymbol{I})$, we can utilize this property to recover $\x$ from $\y$. Mathematically we want to
\begin{equation} \label{eqn:sparserep}
\mathrm{find~} \z , \mathrm{such~that~} \, \abs{\mathrm{supp}(\z)} \ll N\,,\, \norm{\y-\D\z} \leq \epsilon,
\end{equation}
where $\epsilon$ is implicitly related to our observation noise-level $\sigma$. Finding the sparse code (latent representation), $\z$, allows us to reconstruct our signal of
interest through the dictionary via $\hat{\x} = \D\z$.

As stated, the sparse representation problem is intractable for large signals (by its
combinatorial nature). Replacing the condition on the number of non-zero
representation coefficients with the $\ell_1$-norm, we obtain the convex
relaxation known as Basis Pursuit DeNoising (BPDN)
\begin{equation} \label{eqn:bpdn}
\underset{\z}{\mathrm{minimize}} ~ \frac{1}{2}\norm{\y -\D\z}_2^2 + \lambda \norm{\z}_1.
\end{equation}
A simple method for tackling BPDN is the Iterative Shrinkage Thresholding
Algorithm (ISTA) \cite{Beck2009}, whose iterates are defined by a fixed-point condition on the
sub-gradient of \eqref{eqn:bpdn},

\begin{equation} \label{eqn:ista}
\z^{(k+1)} \coloneqq \mathrm{ST}(\z^{(k)} - \eta \D^\top(\D\z^{(k)} - \y), \,
\eta\lambda),
\end{equation}
where $\mathrm{ST}(\x,\tau)[n] \coloneqq \sign(\x[n])\max(0, \abs{\x[n]} - \tau)$
is the element-wise soft-thresholding operator with threshold $\tau \geq 0$, $\x[n]$ denotes element $n$ of $\x$, and
$\eta$ is a step-size parameter. ISTA is attractive for its simplicity of
implementation and low computational cost per-iteration.

It is clear that the success of the spare-representation prior is largely dependent 
on the choice of dictionary $\bm{D}$.
Until recently, the \soa in low-level image
processing tasks was dominated by methods with data-driven dictionaries,
such as K-SVD \cite{KSVD}. These dictionary-learning methods are often formulated
by minimizing the BPDN functional \eqref{eqn:bpdn} over the dictionary $\D$ as
well as sparse-codes $\z$,
\begin{equation} \label{eqn:dict_learn}
\underset{\{\z_i\}, \D \in \mathcal{C}}{\mathrm{minimize}} ~ \sum_{\y_i \in \mathcal{D}} \frac{1}{2}\norm{\y_i -\D\z_i}_2^2 + \lambda \norm{\z_i}_1,
\end{equation}
where we restrict the columns of $\D$ to the constraint set of the $\ell_2$-unit-ball, $\mathcal{C} = \{\D : \norm{\D_{:,m}}_2 \leq 1 \}$, to avoid arbitrary
scaling of coefficients ($\z_i[n]$), and $\mathcal{D}$ denotes the dataset of observations. The resulting non-convex problem is generally solved to a
local optimum via alternating between sparse-coding (ex. via ISTA) and updating the
dictionary (ex. via projected gradient descent) \cite{mairal2009online}.

In the case of a convolutional dictionary with $M$ filters $\bm{d}^m\in \R^{P}$, their integer translates form the columns of $\bm{D}$. The application of the dictionary (synthesis convolution) is given by 
$\bm{D}\z = \sum_{m=1}^M \bm{d}^m \ast \z^m$, where $\z^m$ is the $m$-th subband of $\z$. 
The corresponding analysis convolution is defined subband-wise by $(\bm{D}^\top \x)^m = \overline{\bm{d}^m} \ast \x$ where $\overline{\bm{d}^m}$ denotes the reversal of the filter. In the case of 2D input signals $\x \in \R^{\sqrt{N}\times\sqrt{N}}$\footnote{More generally, we can consider $x\in \R^{N_1\times N_2}$ with $N_1N_2=N$.}, we consider 2D square filters $\bm{d}^m \in \R^{\sqrt{P}\times\sqrt{P}}$.

\subsection{Unrolled Networks} \label{sec:unrolled}

The iterates of ISTA closely resemble that of a DNN.
This connection was formally established in the seminal work of Gregor et al.
\cite{Gregor2010} in which ISTA was truncated to $K$ iterations, and operators $\W_1
= (I - \eta\D^\top\D)$ and $\W_2=\eta \D^\top$ were replaced by learned dense matrices. The
resulting recurrent neural network, known as LISTA, was trained to
approximate sparse-codes from natural image patches given a known dictionary $\D$.
Accelerated convergence of ISTA was empirically shown\cite{Gregor2010}.
More recently, convergence guarantees of the LISTA network under 
different weight parameterizations have been
derived\cite{Chen2018}.

The success of LISTA has since inspired many works, not only in 
sparse-approximation, but as a tool for principled model
construction. In the latter case, classical algorithms such as
gradient-descent, proximal-gradient descent \cite{ongie2020deep}, alternating direction method of multipliers \cite{Yang2020}, and the primal-dual hybrid gradient method \cite{Adler2018} have been used as
starting points for deriving DNN architectures. However, the prevailing approach
in these works is to replace linear-operators of the original algorithm with
non-linear DNN layers (ex.  \texttt{conv} $\rightarrow$ \texttt{batch-norm}
$\rightarrow$ \texttt{relu}) \cite{Adler2018}, or
full-fledged DNNs \cite{ongie2020deep, Zheng2021CVPR} (e.g. UNet \cite{unet}), and sometimes to embed the entire
scheme in a residual learning strategy \cite{Valsesia2020}. This is in contrast to LISTA, which
parameterized the linear-operators of the original ISTA algorithm directly as linear-operators.

These unrolled network approaches, that do not follow a direct parameterization of
their original algorithms, ultimately end up introducing black-box DNNs, and so the question of principled network construction is not fully addressed. As a result, analyzing or improving such unrolled
networks may present challenges inherited from the use of poorly-understood building blocks \cite{santurkar2018does}.

\begin{figure*}[ht]
    \centering
    \includegraphics[width=0.9\linewidth]{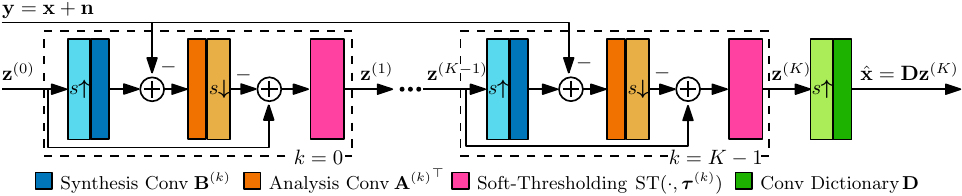}
    \caption{Block diagram of CDLNet. Analysis and synthesis convolutions map from $C\in\{1,3\} \rightarrow M$ and $M \rightarrow C\in\{1,3\}$ channels, respectively. We say that CDLNet does not process signals in a ``learned feature domain" to differentiate from the usage of multi-channel filtering ($M \rightarrow M$ channels) in DNNs such as DnCNN \cite{DnCNN}. Also note that CDLNet does not use batch-normalization or residual learning, in contrast to DnCNN \cite{DnCNN}.}
    \label{fig:block_diag}
\end{figure*}

\subsection{Dictionary Learning Networks} \label{sec:dlnets}
Recently, networks derived from dictionary learning algorithms have gained attention in the image-processing literature precisely for their offer of more principled network construction and high-performance at a relatively low learned-parameter count. In unrolling dictionary learning, ISTA is generally chosen as the classical sparse-pursuit algorithm for its differentiability. The resulting networks have the form of LISTA followed by a learned dictionary $\D$, i.e. a learned sparse coding followed by a synthesis dictionary to bring us from the latent code-domain back to the image domain. Its forward pass is interpreted to perform accelerated (learned) sparse-coding and its backward pass updates the dictionary and (learned sparse-coder), similar to how classical dictionary algorithms iterate between sparse-coding and dictionary updates. The presence of a learned synthesis dictionary $\D$ at the end of the LISTA network allows one to train LISTA in an unsupervised, ``task-driven" manner, i.e. tailored to the loss function and image-processing problem at hand.

Sreter et al. \cite{Sreter2018} proposed such a convolutional dictionary learning (CDL) network (Conv-LISTA $\rightarrow$ Conv-$\D$) for the tasks of denoising and inpainting, known as Approximate Convolutional Sparse Coding (ACSC). They demonstrated results favorable to classical dictionary learning methods such as K-SVD, however, not on par with the \soa DNNs. Simon et al. \cite{Simon2019} attributed ACSC's lack of empirical success (against DNNs) to the supposed ill-conditioning of the convolutional dictionary model. Their argument led them to propose a CDL network similar to that of ACSC, however, with large-strided convolution operators (stride on the order of the size of their filters). Additionally, their proposed network (CSCNet) processes all possible shifts of the input image and averages the output (``shift-averaging"), moving their architecture towards a patch-based dictionary-learning approach such as K-SVD. CSCNet is able to compete well with popular convolutional neural networks (CNNs) without non-local operations. Our proposed method is also derived from an ISTA-CDL algorithm, but instead unties the weights along iterations. With this alternative parameterization, we use small-strided convolution operations as a tool to manage computational complexity. We also construct our network to be noise-level adaptive, whereas CSCNet does not.

Lecouat et al. \cite{Lecouat2020nonLocal} later proposed a network derived from a differentiable patch-based dictionary learning algorithm with non-local self-similarity. They demonstrate competitive performance to \soa DNNs (with non-local operations) on denoising and demosaicing tasks separately. In contrast, our work relies only on convolution operations and considers the task of joint denoising and demosaicing with a single network. Lecouat et al. \cite{Lecouat2020nonLocal} also propose a noise-adaptive scheme for noise-level blind image denoising. However, their scheme only considers a discrete set of noise-levels, whereas our scheme introduces a parameterization that allows for a continuous range of input noise-levels.

Similarly, Scetbon et al. \cite{Scetbon2021} constructs a patch-processsing based unrolled dictionary-learning architecture (LKSVD). LKSVD also considers leveraging the sparse-coding interpretation of the network by adapting their Lagrange multiplier term ($\lambda$) to the input image/patch. However, LKSVD does this by learning a black-box DNN mapping patches to $\lambda$. In CDLNet, we explicitly model an affine relationship between the input noise-level and thresholds of the network. We demonstrate that this paramterization allows for near-perfect generalization to input noise-levels unseen during training. The subject of noise-level generalization is not investigated any of the above mentioned works \cite{Sreter2018,Simon2019,Lecouat2020nonLocal, Scetbon2021}.

\subsection{Popular CNNs and Noise-Adaptivity} \label{sec:popularnoise}
DnCNN \cite{DnCNN} continues to be used as a measuring stick in the image-processing literature, despite higher-performing networks since being published \cite{liu2018non,Lecouat2020nonLocal,Zheng2021CVPR}. These more recent networks often achieve their superior performance via more sophisticated architectures (such as those using non-local operations). In this work, our primary focus is on a novel formulation of noise-adaptivity in DNNs. Though this adaptivity is not limited to application in fully convolutional neural networks (FCNNs), our proposed network is fully convolutional. Thus, we limit our numerical comparisons to other FCNNs.

DnCNN, and many architectures since, choose to handle varying input noise-levels by presenting samples with varying noise-levels during training. While this generally yields only a small performance drop compared to single noise-level trained models within the training range, catastrophic failure may be observed when tested outside their training noise-level range \cite{Mohan2020}. Mohan et al. \cite{Mohan2020} recently investigated this phenomenon and proposed the removal of the network's bias parameters as an effective remedy. Their corresponding bias-free DnCNN architecture (BFCNN) is able to generalize its denoising performance outside the training range, although with a further drop in denoising performance inside the training range. In CDLNet we do not employ bias weights in our convolution operators, however, our learned soft-thresholding effectively introduces a bias term that can be modeled as a function of the input noise-level. Through this explicitly defined noise-level adaptivity, we demonstrate that denoising generalization, such as obtained by BFCNN \cite{Mohan2020}, may be instead obtained through principled network construction and show a smaller drop in denoising performance as compared to BFCNN \cite{Mohan2020} within the network's training range. Furthermore, we show the first instance of similar generalization capabilities under additional tasks such as color denoising, joint denoising and demosaicing, and unsupervised learning.

FFDNet \cite{FFDNet} represents another common approach to noise-level adaptivity in the literature. In this architecture, the input noise-level is concatenated to input image before being passed to the network. The network is also trained over a wide range of noise-levels. This method treats the network as a black-box, simply providing another piece of information as input. In contrast, the proposed CDLNet augments the thresholds of the network in each layer, explicitly modeling the functional form of noise-adaptivity and allowing superior generalization outside the training range of noise-levels.

\section{Proposed Method}
\label{sec:ProposedMethod}

\subsection{CDLNet Architecture for Denoising} 
\label{sec:architecture}

In this section we introduce the convolutional dictionary learning network (CDLNet) for natural-image denoising. Let $\y = \x + \bnu$ be a noisy observed image with $\x$ and $\bnu$ denoting the corresponding clean image and the noise vector, respectively. To derive the CDLNet architecture, we parameterize $K$ ISTA iterations \eqref{eqn:ista} using learnable convolutional analysis (${\bm{A}^{(k)}}^\top$) and synthesis ($\bm{B}^{(k)}$) operators. The CDLNet architecture thus involves a convolutional LISTA encoder followed by a learned convolutional synthesis dictionary, $\bm{D}$,
\begin{gather} \label{eqn:CDLNet}
\hat{\x} = \bm{D}\z^{(K)}, \quad \z^{(0)} = \bm{0}, \quad k=0,1,\dots, K-1,\\
\z^{(k+1)} = \ST\left(\z^{(k)} - {\bm{A}^{(k)}}^\top (\bm{B}^{(k)}\z^{(k)} - \y), \, \boldsymbol{\tau}^{(k)} \right), \nonumber
\end{gather}
where $\Theta = \{[\bm{A}^{(k)}, \bm{B}^{(k)}, \boldsymbol{\tau}^{(k)}]_{k=0}^{K-1}, \bm{D}\}$ are the set of learned parameters. For $M$-subband dictionaries, $\bm{B}^{(k)}$ and $\bm{D}$ are $M$-subband convolutional synthesis operators, and ${\bm{A}^{(k)}}^\top$ are $M$-subband convolutional analysis operators (with $C=1$ or $3$ channels in each subband filter for grayscale or color images, respectively). The non-negative thresholds $\boldsymbol{\tau}^{(k)} \in \R^M$ are subband dependent, corresponding to a learned weight for the $\ell_1$-norm in \eqref{eqn:dict_learn}. The output denoised image is given by $\hat{\x} = f_\Theta(\y) = \bm{D}\z^{(K)}$. The block diagram of the proposed CDLNet is given in Fig.~\ref{fig:block_diag}.

The convolution analysis and synthesis operators as shown are highly redundant transformations, increasing the number of elements in the signal domain from $CN$ (for $\x \in \R^{\sqrt{N}\times\sqrt{N}\times C}$) to $MN$ in the transform domain. This level of redundancy can become computationally burdensome as the number of filters ($M$) increases. To mitigate this, as shown in \Fig \ref{fig:block_diag}, we can add a sub-sampling operator with stride $s$ in both horizontal and vertical directions after each analysis operator and a corresponding zero-filling operator with the same stride before the synthesis operator. The computational complexity of this network is roughly $O(KMCNP/s^2)$ for input signals in $\R^{\sqrt{N}\times\sqrt{N}\times C}$ and filters of size $\sqrt{P}\times\sqrt{P}$. This is equivalent to replacing each of the convolution operators ${\bm{A}^{(k)}}^\top,\bm{B}^{(k)},~\text{or}~\bm{D}$ by ${\bm{A}^{(k)}}^\top\Delta_s^\top ,\Delta_s\bm{B}^{(k)},~\text{or}~\Delta_s\bm{D}$, where $\Delta_s$ is the zero-filling operator with stride $s$ and $\Delta_s^\top$ is the sub-sampling operator with stride~$s$. We have found that  adding such sub-sampling operators allows for the possibility of learning more diverse filters while keeping the computation under control. The stride~$s$, the number of subbands~$M$, and number of iterations~$K$ can be tuned to achieve the desired trade-off between the complexity and denoising quality (see Table \ref{tab:depth_width} in Appendix).

\subsection{CDLNet Architecture for Demosaicing}

The demosaicing problem aims to fill-in the unmeasured color values from a color filter array. For a colored image $\x\in\R^{\sqrt{N}\times\sqrt{N}\times 3}$, let $\m \in \{0,1\}^{\sqrt{N}\times \sqrt{N} \times 3}$ denote the mask signal with an element equal to $1$ for color pixels that are measured, and $0$ otherwise. Then, the measured signal can be modeled as $\y= \m \circ(\x+\bnu)$ where $\bnu \sim\mathcal{N}(0,\sigma^2\bm{I})$. Consequently, the data-fidelity term in \eqref{eqn:dict_learn} has to change such that it only considers the observed values. Note that the AWGN model is applicable in the case of gamma-corrected and white-balanced images, as is the case in this work \cite{Gharbi2016}. Mathematically, our dictionary learning problem becomes,
\begin{equation} \label{eqn:dmsc_dict_learn}
\underset{\bm{D}\in \mathcal{C}, \{\z_i\}}{\mathrm{minimize}} ~  \sum_{i} \frac{1}{2}\norm{\m \circ \bm{D}\z_i -\y_i}_2^2 + \lambda \norm{\z_i}_1.
\end{equation}
The corresponding ISTA iteration can be written as %
\begin{equation} 
\z^{(k+1)} = \ST\left(\z^{(k)} - 
\eta^{(k)}\bm{D}^\top  (\m \circ\bm{D}\z^{(k)} - \y),\, \eta^{(k)}\lambda \right).
\end{equation}
The masking operator $(\m \circ \cdot)$ will be added to each layer of the network in the CDLNet architecture,
\begin{gather} \label{eqn:DMSC_CDLNet}
\hat{\x} = \bm{D}\z^{(K)}, \quad \z^{(0)} = \bm{0}, \quad k=0,1,\dots, K-1,\\
\z^{(k+1)} = \ST\left(\z^{(k)} - {\bm{A}^{(k)}}^\top (\m \circ \bm{B}^{(k)}\z^{(k)} - \y), \, \boldsymbol{\tau}^{(k)} \right). \nonumber
\end{gather}
\subsection{Noise-Adaptive Thresholds} 
\label{sec:architectureNoise}

We leverage CDLNet's signal-processing and optimization derivation to provide adaptation to varying input noise-levels in a single model. Given that the noise standard deviation $\sigma$ can be estimated from the input image \cite{Chang2000,Liu2013}, we parameterize the learned thresholds in \eqref{eqn:CDLNet} as an affine function of the estimated input noise standard deviation ($\sigma$), 
\begin{equation} \label{eqn:thresh}
\boldsymbol{\tau}^{(k)} = \boldsymbol{\tau}^{(k)}_0 + \boldsymbol{\tau}^{(k)}_1 \sigma,\quad \mathrm{such~that} \quad \boldsymbol{\tau}^{(k)}_{\{0,1\}} \geq0,
\end{equation}
for $k=0,1,\dots, K-1$ where $\boldsymbol{\tau}^{(k)}_{\{0,1\}}\in\R^M$ are learned parameters. We propose this parameterization of the thresholds based on the ``universal threshold theorem" \cite{Mallat}, from the Wavelet denoising literature,
$\tau = \sigma \sqrt{2\log_e N}$.
This can be shown to minimize the upper bound of the risk function of an element-wise denoising operator \cite{Mallat}. In order to avoid losing the non-linear operation in CDLNet for small values of $\sigma$, we prescribe an affine model for the adaptive thresholds. 

With this parameterization, we interpret the proportionality constant $\boldsymbol{\tau}^{(k)}_1[m]$ as learning the gain factor between the image domain noise-level $\sigma$ and the $m$-th subband's noise-level at layer $k$. This framework has the added benefit of decoupling noise-level estimation from denoising, allowing for trade-off between accurate estimation and speed at inference time. We explore this trade-off using two different noise-level estimation algorithms at inference time in Section \ref{sec:exp_blind}. 

\subsection{Relation to Popular DNNs} \label{sec:dnns}
In this section we highlight some of the similarities and differences between CDLNet and popular black-box DNNs.

\textbf{Thresholding}: Current popular image-restoration networks employ the point-wise non-linearity of the rectified linear unit (ReLU). Combined with a bias parameter from either batch-normalization or the convolution layer, the ReLU can be viewed as a thresholding operator for producing non-negative coefficients. To see this, consider a convolution operator $\bm{C}$ and a channel-wise bias vector $\bm{b}$ with signal $\bm{x} \in \R^N$. A standard layer is then expressed as follows,
\begin{align*}
\mathrm{ReLU}(\bm{C}\x + \bm{b})[n] &= \max(0, \z[n] + \bm{b}[n]) \\ &= \begin{cases} \z[n] + \bm{b}[n], & \z[n] > -\bm{b}[n] \\ 0,& \mathrm{otherwise} \end{cases},
\end{align*}
where $\z = \bm{C}\x$. Hence the bias$+$ReLU acts as a non-negative thresholding with threshold $-\bm{b}$. This thresholding shrinks the coefficient magnitude given a negative bias term and increases it given a positive bias term. 

In contrast, the soft-thresholding of CDLNet is a shrinkage-thresholding which produces both positive and negative coefficients. This points to a possible source of reduced redundancy as compared to ReLU-based networks, which have been empirically observed to learn filter pairs of opposite signs in their convolution layers \cite{Shang2016}.

\textbf{Feature-Domain Processing}: 
CDLNet differs from DnCNN and other popular DNN architectures because it does not employ inter-subband convolution operators. Visualization and understanding of the network's learned operators and representations is often a non-trivial task for black box models. This is in part due to the use of inter-subband ($M\rightarrow M$) convolution operators whose multi-channel filters cannot be easily visualized. In contrast, CDLNet's analysis and synthesis filters consist of $C$-channel filters ($C\in\{1,3\}$) and may be readily shown (see Section \ref{sec:results}). This allows for greater interpretability of the network's intermediate representations.

Feature domain processing is often motivated by intuition borrowed from the neuroscience literature, wherein layered representation is beneficial for obtaining semantic meaning from images \cite{lindsay2021convolutional}. However, for low-level image processing tasks such as denoising and demosaicing, it is not clear that such a semantic understanding of the input would be beneficial. Indeed we find that CDLNet's subband representation alone is enough to achieve performance on-par with \soa FCNNs in denoising and demosaicing tasks.

\section{Experimental results} \label{sec:results}
\subsection{Supervised and Unsupervised Training of CDLNet}
In this section we discuss the training loss functions that can be utilized to train CDLNet. In the case where a dataset, $\mathcal{D}$, of corresponding noisy and clean signals is available,  $\{\y_i,\x_i\} \in \mathcal{D}$, we can train the network using the mean square error (MSE) loss function. Let $f_{\Theta}(\cdot)$ denote the CDLNet model. Then the supervised loss function can be written as 
\begin{equation} \label{eqn:mse}
\underset{
\substack{
\bm{A}^{(k)} \in \mathcal{C}, ~ \bm{B}^{(k)} \in \mathcal{C} \\\bm{D} \in \mathcal{C}, ~
\boldsymbol{\tau}^{(k)} \geq 0
}
}{\mathrm{minimize}} \quad \sum_{\{\y_i,\x_i\} \in \mathcal{D}} \|\x_i - f_{\Theta}(\y_i)\|_2^2,
\end{equation}
where $\mathcal{C} = \{\D : \norm{\D_{:,m}}_2 \leq 1 \}$ is the unit-ball constraint and is imposed on all convolution operators.

In the case where corresponding clean signals are not available, we can use Stein's unbiased risk estimator (SURE) \cite{stein1981estimation} as an unbiased estimate of the MSE \cite{metzler2018unsupervised}. We have 
\begin{equation}
    \begin{aligned}
&\mathbb{E}\left[\|\x - f_{\Theta}(\y)\|_2^2\right] =\\& \mathbb{E} \left[\|\y - f_{\Theta}(\y)\|_2^2\right] - N \sigma^2 + 2\sigma^2 \divrg_{\y} \left(f_{\Theta}(\y)\right)
\end{aligned}
\label{Eq:SURELoss}
\end{equation}
where $\divrg(\cdot)$ is the divergence and is defined as 
\begin{equation}
\divrg_{\y} \left(f_{\Theta}(\y)\right) = \sum_n \frac{\partial f_{\Theta}(\y)[n]}{\partial\y[n]}.
\end{equation}

The challenge in using the SURE loss \eqref{Eq:SURELoss} is calculating the divergence term. Following \cite{ramani2008monte}, we use a Monte Carlo method. Let $\bm{b}$ be an i.i.d Gaussian distributed random vector with unit variance elements. Then, the divergence term can be estimated as 
\begin{equation}
\divrg_{\y} \left(f_{\Theta}(\y)\right) \approx \bm{b}^T \left(\frac{f_{\Theta}(\y+h\bm{b})-f_{\Theta}(\y)}{h}\right)
\end{equation}
where we set $h$ to 1e-3. The combination of this estimate with the SURE loss \eqref{Eq:SURELoss} enables us to train CDLNet without use of ground truth data. Note that the computation of the divergence term requires two forward passes through the network and as a result training with this method is more computationally demanding. Since the divergence term is calculated with a Monte Carlo sampling, it is common to refer to the obtained loss function as MC-SURE \cite{metzler2018unsupervised}.

\subsection{Training and Inference Setup}
\textbf{Architectures}:  The name CDLNet refers to models with adaptive thresholds trained over a noise range. We experiment with both low and high parameter count CDLNet models. We denote models with a low parameter count via the prefix ``small-". Models trained on a single noise-level are denoted via the suffix ``-S", and models trained over a noise range but without adaptive thresholds via the suffix ``-B". The prefix ``JDD" is given to models trained for the joint denoising and demosaicing task, and the same naming convention mentioned previously also applies. Unless otherwise specified, Table \ref{tab:arch} lists the architectures of each model.

\begin{table}[tb]
\caption{Architectures of the CDLNet models and variants presented in the experimental section. 
We use $C=3$ for color denoising and JDD networks and $C=1$ for grayscale networks.
A filter size of $P = 7\times 7$ is used for all models. The listed stride $s$ is used unless otherwise specified.}
\centering
\resizebox{0.85\linewidth}{!}{%
\begin{tabular}{cccccc} \hline
     Name & Task & $K$ & $M$ & $s$ & params \\ \hline
     small-CDLNet-S & Gray & 20 & 32 & 1 & 64k\\ 
     small-CDLNet-S & Color & 20 & 32 & 1 & 190k\\ 
     CDLNet(-S,-B) & Gray & 30 & 169 & 2 & 507k \\ 
     CDLNet(-S,-B) & Color & 22 & 100 & 2 & 652k \\ 
     JDD small-CDLNet & JDD & 30 & 42 & 1 & 373k \\ 
     JDD CDLNet(-S,-B) & JDD & 42 & 64 & 1 & 796k \\ \hline
\end{tabular}
}
\label{tab:arch}
\end{table}

\textbf{Dataset}: CDLNet and all its variants are trained on the BSD432 and CBSD432 dataset \cite{bsd} for grayscale and color image inputs, respectively. Note that the JDD CDLNet models are also trained on the CBSD432 dataset with the RGGB Bayer mask \cite{Gharbi2016} synthetically applied. Input signals are pre-processed with division by $255$, and per-image mean-subtraction. Training inputs are additionally pre-processed by random crops of $128\times 128$, random flips, and random rotations. Models trained across noise-levels are done so by uniform sampling of $\sigma \in \sigma^{\mathrm{train}}$. Noise-level $\sigma$ is listed relative to signal intensity on a range of $[0,255]$.

\textbf{Training}: Models are trained by the ADAM \cite{adam} optimizer on the MSE loss \eqref{eqn:mse}, or the MC-SURE loss \eqref{Eq:SURELoss} when specified. The unit-ball constraint $\mathcal{C} = \{\D : \norm{\D_{:,m}}_2 \leq 1 \}$ is imposed on all convolution operators. Additionally, we require $\boldsymbol{\tau}^{(k)}_{\{0,1\}}\geq 0$. Constraints are enforced by projection after each gradient descent step. A mini-batch size of $10$ samples is used. An initial learning rate of 1e-3 is used, and reduced by a factor of $0.95$ every $50$ epochs for a maximum of $6000$ epochs or until convergence. Similar to the method in \cite{Lecouat2020Games}, we backtrack our model to the nearest checkpoint upon divergence of PSNR on the Kodak dataset \cite{Kodak}, reducing the learning rate by a factor of $0.8$. Ground-truth noise-level is given to models during training.

\textbf{Initialization}: We initialize all synthesis convolution operators with the same filterbank $\{\bm{w}^m\}_{m=1}^M$ drawn from independent standard normal distributions, and all analysis convolution operators with the transposed weights (flipped filters) $\{\overline{\bm{w}^m}\}_{m=1}^M$. This setup will initialize the network as $K$ ISTA iterations. Following \cite{Simon2019}, as an initialization step, we normalize the convolution operators by their spectral norm, in correspondence with the maximum uniform step-size of ISTA. Furthermore, the use of standard normal distribution comes from the intuition that the majority of our learned subbands will consist of band-pass signals modeling image texture and so our filters should be zero-mean. 

\textbf{Noise-level estimation}: We look at employing two different noise-level estimation algorithms for blind denoising: one based on the median absolute deviation (MAD) of the input's diagonal wavelet coefficients \cite{Chang2000}, and the other based on the principal component analysis (PCA) of a subset of the input's patches \cite{Liu2013}. These estimators offer two ends of the trade-off between speed (MAD) and accuracy (PCA). However, unless otherwise noted the ground truth noise-level is given.

\textbf{Hardware}: Experiments were conducted with an Intel Xeon(R) Platinum 8268 CPU at 2.90GHz, an Nvidia RTX 8000 GPU, and 4GB of RAM, running Linux version 3.10.0. Training times for CDLNet models on this setup are between 9-24 hours, depending on network complexity. Code is available at \url{https://github.com/nikopj/CDLNet-OJSP}.

\subsection{Single Noise-level Denoising}
In this section, we present comparison results for cases where the networks are trained for a single noise-level and tested on the same noise-level.

\textbf{Grayscale denoising:} Table \ref{table:graysingle} shows the results of the CDLNet model in the small parameter count regime  (small-CDLNet-S) against the other leading CDL based DNN, CSCNet \cite{Simon2019}, and in the large parameter count regime (CDLNet-S) against \soa fully convolutional denoising networks, DnCNN \cite{DnCNN} and FFDNet \cite{FFDNet}. Small-CDLNet-S performs competitively with  CSCNet \cite{Simon2019} in an order of magnitude less computational time at inference by forgoing ``shift-averaging",
using fewer filters, and untying weights between unrollings. CDLNet-S outperforms DnCNN \cite{DnCNN} and FFDNet \cite{FFDNet} by scaling its parameter count to a comparable size. Inference timings given show that our larger model's use of stride yields a manageable computational complexity. The non-learned method BM3D \cite{bm3d} is given as a classical baseline, and its timing is an order of magnitude greater than the learned methods (and thus omitted from the table). 
\begin{table}[ht]
\centering
\caption{Grayscale image denoising performance (PSNR) on BSD68 testset ($\sigma = \sigma^{\mathrm{train}} = \sigma^{\mathrm{test}}$), and single image inference time ($\mathrm{ms}$). All learned models trained on BSD432\cite{bsd}, except CSCNet (BSD432\cite{bsd} $+$ Waterloo ED \cite{ma2017waterloo}). PSNRs reported in respective citations.}
\resizebox{\linewidth}{!}{%
\begin{tabular}{cccccccc} \hline
\multirow{2}{*}{Model} & \multirow{2}{*}{Params} & \multicolumn{3}{c}{Noise-level ($\sigma$)} & \multirow{2}{*}{GPU time}\\
 & & 15 & 25 & 50 & \\ \hline
BM3D \cite{bm3d}        &  -  & 31.07 & 28.57 & 25.62 & - \\
CSCNet \cite{Simon2019} & 64k & 31.57 & 29.11 & 26.24 & 143 ms \\
small-CDLNet-S          & 64k & 31.60 & 29.11 & 26.19 & 9 ms \\
FFDNet \cite{FFDNet}    & 486k& 31.63 & 29.19 & \underline{26.29} & 7 ms \\
DnCNN \cite{DnCNN}      & 556k& \underline{31.72}& \underline{29.22} & {26.23} & 23 ms \\
CDLNet-S                & 507k& \bf{31.74} & \bf{29.26} & \bf{26.35} & 15 ms \\\hline
\end{tabular}
}
\label{table:graysingle}
\end{table}

In Table \ref{table:graysingle}, we use stride 1 convolutions in our small model and stride 2 convolutions for our larger model. Table \ref{table:graystride} empirically verifies these choices as optimal by showing the effect of stride on output PSNR, averaged over a grayscale version of the Kodak dataset \cite{Kodak}. We see that the redundancy of the larger model allows for use of stride 2 without a denoising performance penalty. Note that using a larger stride $s$ does not reduce the model parameter count, but reduces the computational complexity by a factor of $s^2$.

\begin{table}[ht]
\centering
\caption{Effect of stride for $\sigma^{\mathrm{train}}=\sigma^{\mathrm{test}}=25$.\\ PSNR values averaged over grayscale Kodak \cite{Kodak} dataset.}
\resizebox{0.8\linewidth}{!}{%
\begin{tabular}{ccccc}
\hline
Stride & 1 & 2 & 4 & 6 \\ \hline
small-CDLNet-S & {\bf 30.19} & 30.09 & 29.75 & 29.21 \\
CDLNet-S & 30.37 & {\bf 30.39} & 30.28 & 29.83 \\ \hline
\end{tabular}%
}
\label{table:graystride}
\end{table}

\textbf{Color denoising:} Table \ref{table:colorsingle} shows the results of the CDLNet model in the small parameter count regime  (small-CDLNet-S) against other leading CDL based DNN, CSCNet \cite{Simon2019}, and in the large parameter count regime (CDLNet-S) against \soa color denoising models DnCNN \cite{DnCNN} and FFDNet \cite{FFDNet}. We observe that our larger model (CDLNet-S) is able to compete well with DnCNN \cite{DnCNN}, and that our smaller model (small-CDLNet-S) not only outperforms CSCNet \cite{Simon2019}, but also FFDNet \cite{FFDNet}. Similarly to the grayscale case, the effect of stride on our color denoising models is shown in Table \ref{table:colorstride}. The performance of the larger model with single stride is omitted due to its excessive computational resource requirements.
 
\begin{table}[ht]
\centering
\caption{Color image denoising performance (PSNR) on CBSD68 testset ($\sigma = \sigma^{\mathrm{train}} = \sigma^{\mathrm{test}}$). All learned models trained on CBSD432\cite{bsd}, except CSCNet (CBSD432\cite{bsd} $+$ Waterloo ED \cite{ma2017waterloo}). PSNRs reported in respective citations.}
\resizebox{\linewidth}{!}{%
\begin{tabular}{cccccccc} \hline
\multirow{2}{*}{Model} & \multirow{2}{*}{Params} & \multicolumn{4}{c}{Noise-level ($\sigma$)} \\
 & & 5 & 15 & 25 & 50 \\ \hline
CBM3D  &  -                    & 40.24 & 33.49 & 30.68 & 27.36 \\
CSCNet \cite{Simon2019} & 186k & -     & 33.83 & 31.18 & 28.00 \\
small-CDLNet-S          & 190k & 40.46 & 33.96 & 31.30 & \underline{28.05} \\
FFDNet \cite{FFDNet}    & 486k & -     & 33.87 & 31.21 & 27.96 \\
DnCNN \cite{DnCNN}      & 668k & \bf{40.50} & \underline{33.99} & \underline{31.31} & 28.01 \\
CDLNet-S                & 652k & \underline{40.48} & \bf{34.03} & \bf{31.37} & \bf 28.15 \\\hline
\end{tabular}
}
\label{table:colorsingle}
\end{table}
\vspace*{-10pt}
\begin{table}[ht]
\centering
\caption{Effect of stride for color image denoising models ($\sigma^{\mathrm{train}}=\sigma^{\mathrm{test}}=25$).\\ PSNR values averaged over Kodak \cite{Kodak} dataset.}
\resizebox{0.65\linewidth}{!}{%
\begin{tabular}{cccc}
\hline
Stride & 1 & 2 & 4 \\ \hline
small-CDLNet-S & {\bf 32.16} & 31.98 & 31.64 \\
CDLNet-S & - & \bf 32.25 & 31.94 \\ \hline
\end{tabular}%
}
\label{table:colorstride}
\end{table}
\vspace*{-10pt}
\subsection{Blind Denoising and Generalization} \label{sec:exp_blind}

\begin{figure}[t] 
    \centering
    \subfloat[PSNR plot $\sigma^{\mathrm{train}}={[01,20]}$ \label{fig:GrayBlindPlot0120}]{%
        \includegraphics[width=0.49\linewidth]{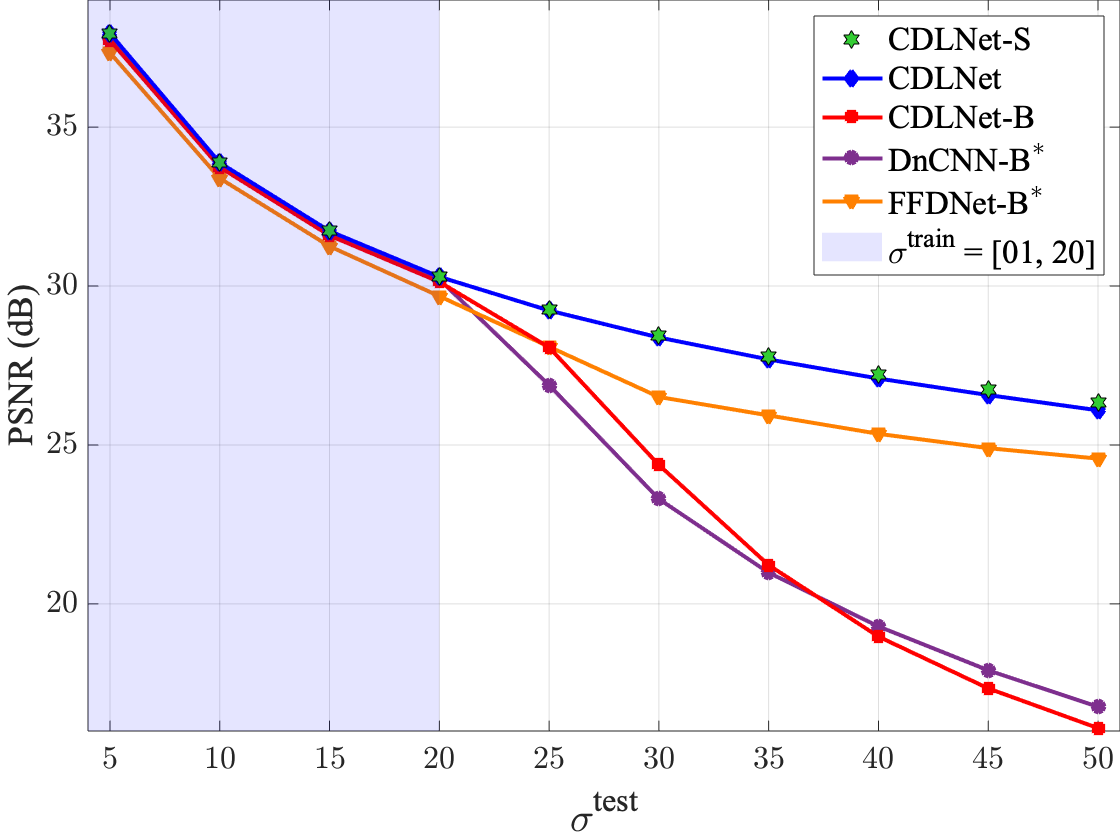}}
  \subfloat[PSNR plot $\sigma^{\mathrm{train}}={[20,30]}$ \label{fig:GrayBlindPlot2030}]{%
        \includegraphics[width=0.49\linewidth]{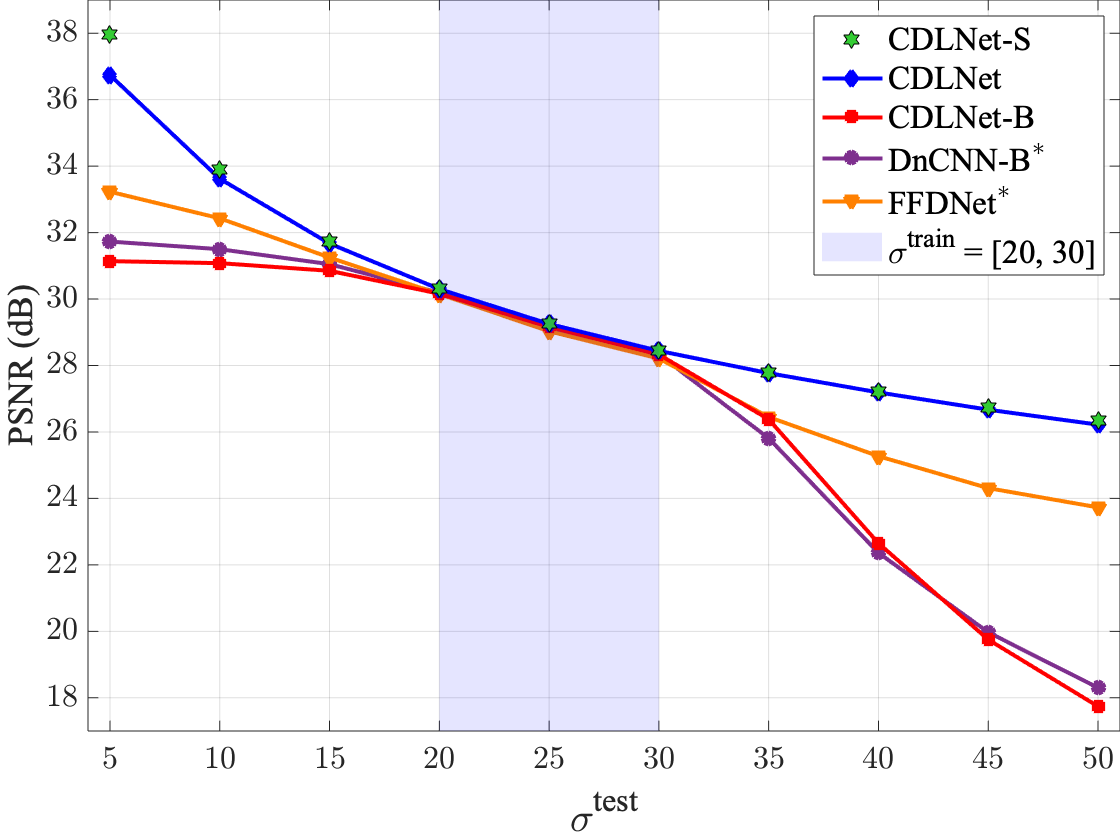}}
    \\\vspace*{8pt}
     \includegraphics[width=\linewidth]{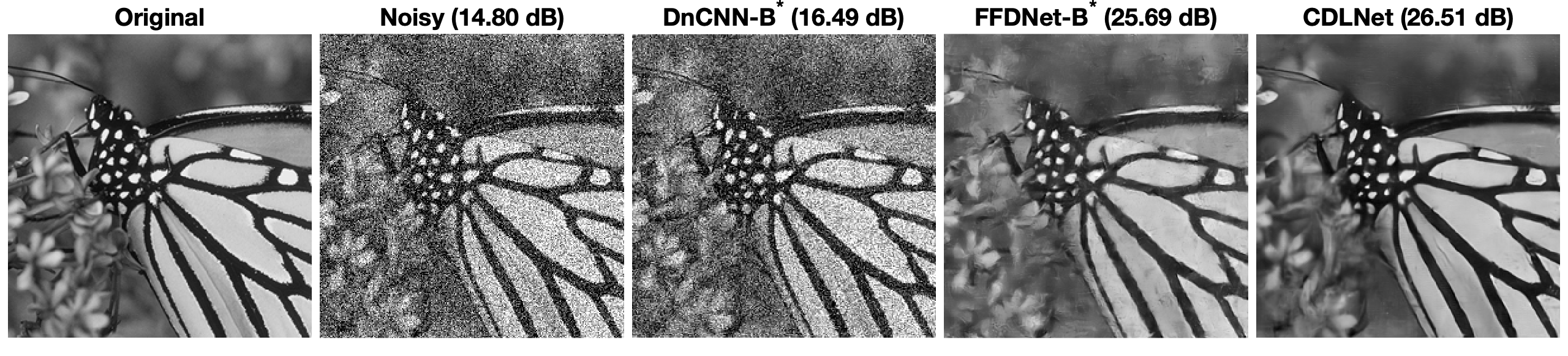}
    \\\vspace*{-10pt}
    \subfloat[Visual comparison for $\sigma^{\mathrm{train}}={[01,20]}$ and $\sigma^{\mathrm{test}}={50}$ \label{fig:ImgGray0120}]{%
       \includegraphics[width=\linewidth]{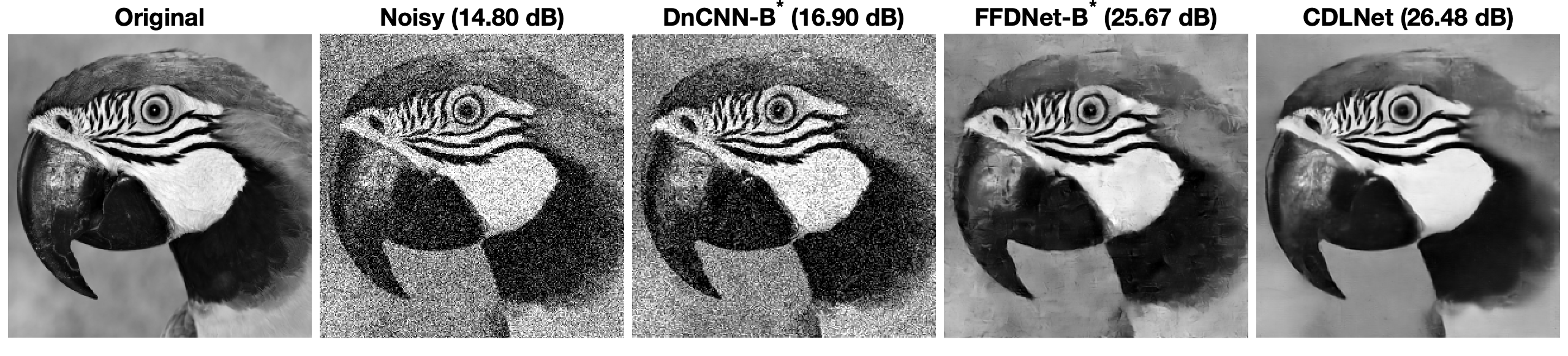}}
    \\\vspace*{8pt}
    \includegraphics[width=\linewidth]{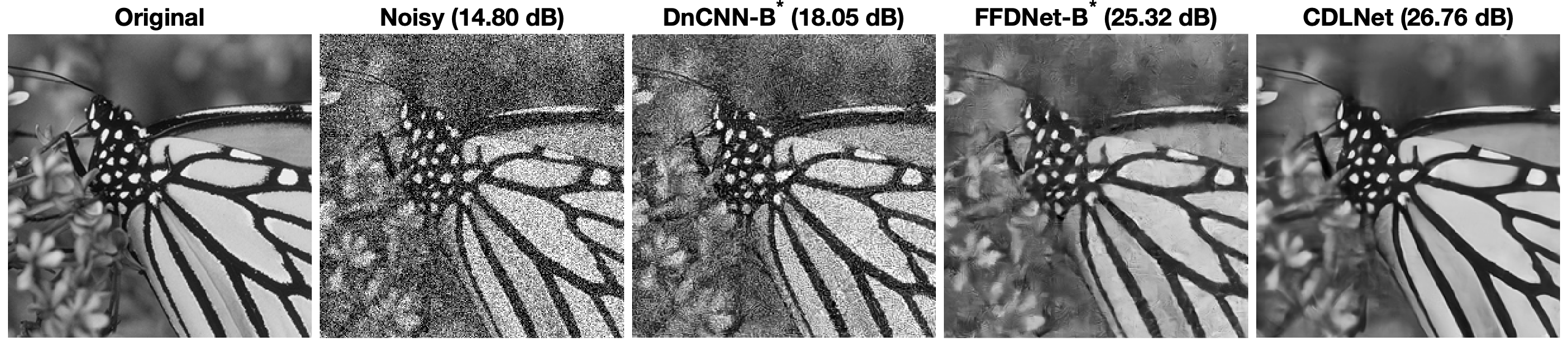}
    \\\vspace*{-10pt}
    \subfloat[Visual comparison for $\sigma^{\mathrm{train}}={[20,30]}$ and $\sigma^{\mathrm{test}}={50}$ \label{fig:ImgGray2030} ]{%
       \includegraphics[width=\linewidth]{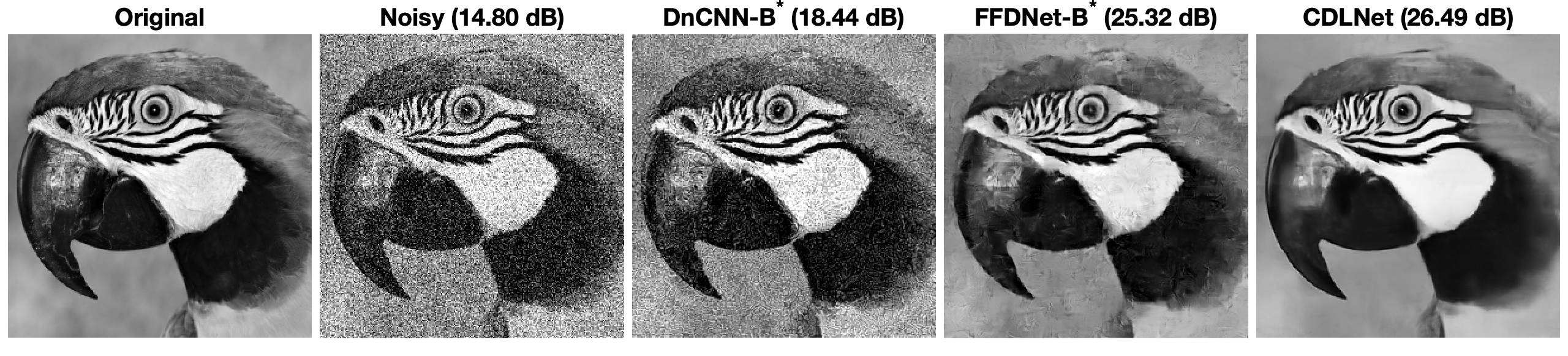}}
    \\\vspace*{8pt}
     \subfloat[PSNR plot $\sigma^{\mathrm{train}}={[01,20]}$ \label{fig:GrayBlindPlot0120Small}]{%
     \includegraphics[width=0.49\linewidth]{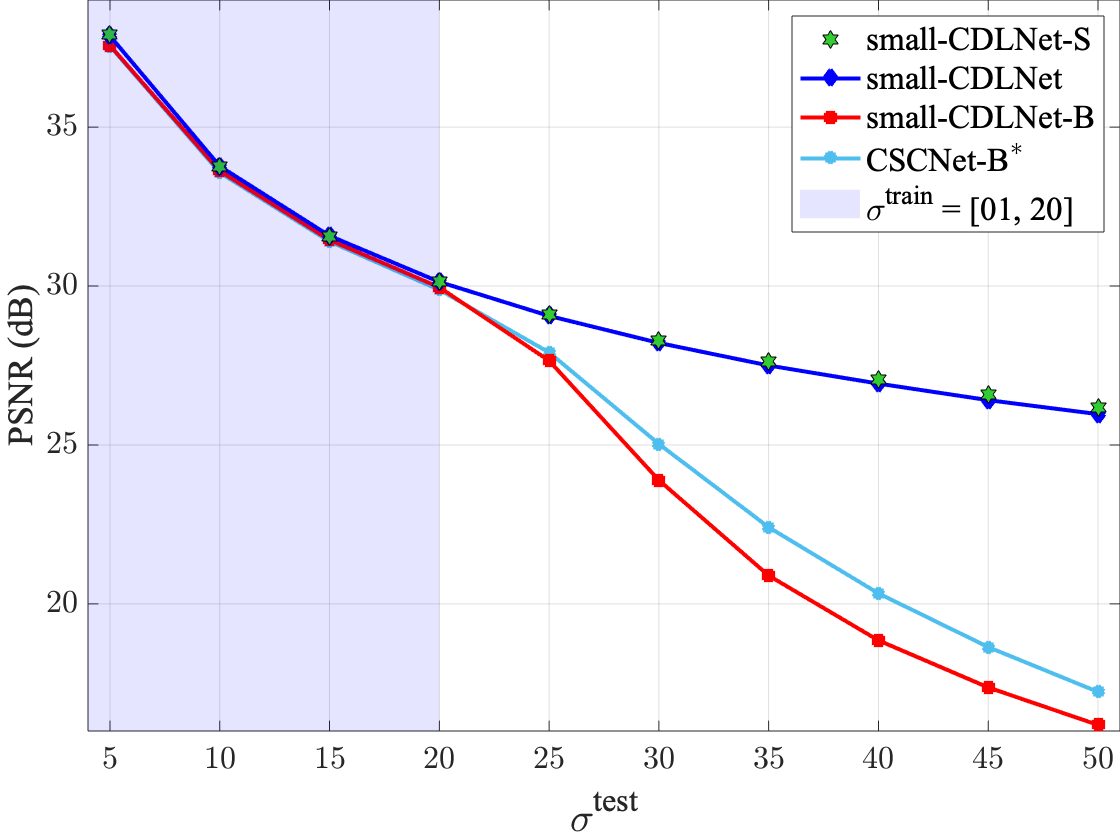}}
    \subfloat[PSNR plot $\sigma^{\mathrm{train}}={[20,30]}$ \label{fig:GrayBlindPlot2030Small}]{%
        \includegraphics[width=0.49\linewidth]{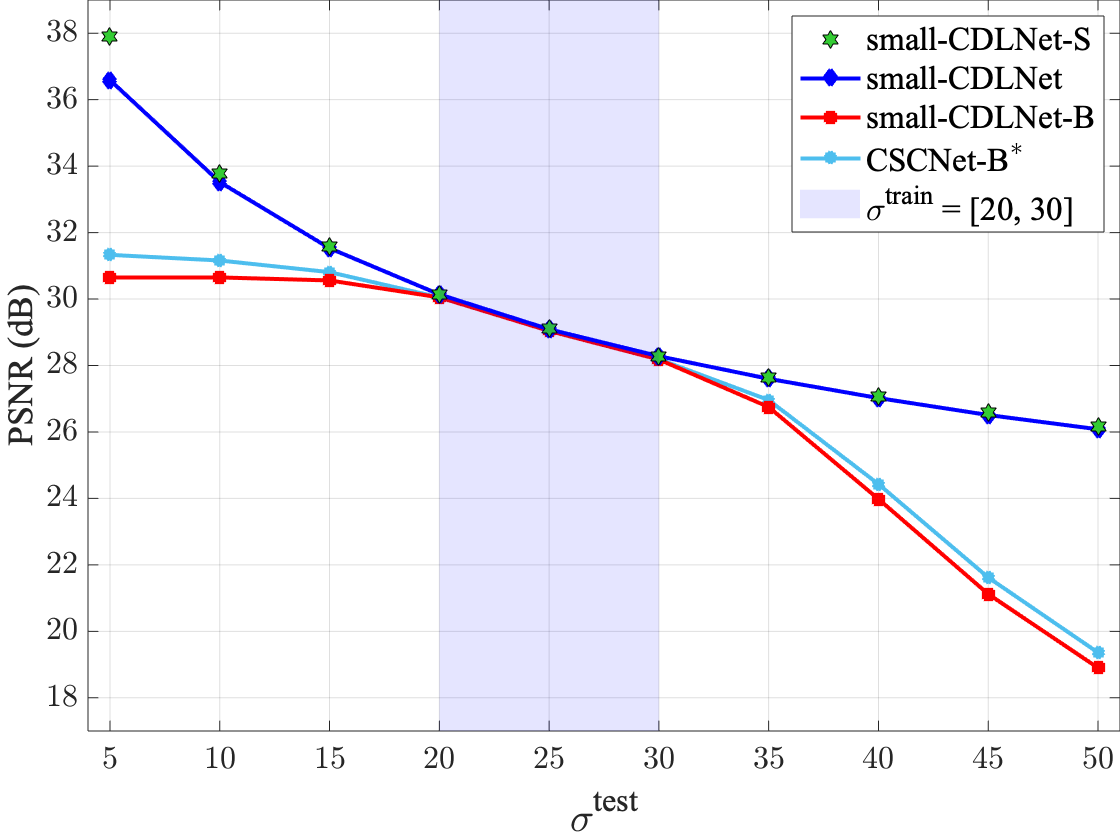}}
  \caption{Performance of different grayscale denoising networks with large parameter count (a,b) and small parameter count (e,f) trained on $\sigma^{\mathrm{train}}$ and tested on different $\sigma^{\mathrm{test}}$. Average PSNR calculated over BSD68 \cite{bsd}. (c,d) Visual comparison of different networks tested on noise-level $\sigma^{\mathrm{test}}=50$. Details are better visible by zooming.}
  \label{fig:ImgGray} 
\end{figure}

\begin{figure}[t] 
    \centering
    \subfloat[PSNR plot $\sigma^{\mathrm{train}}={[01,20]}$ \label{fig:ColorBlindPlot0120}]{%
        \includegraphics[width=0.49\linewidth]{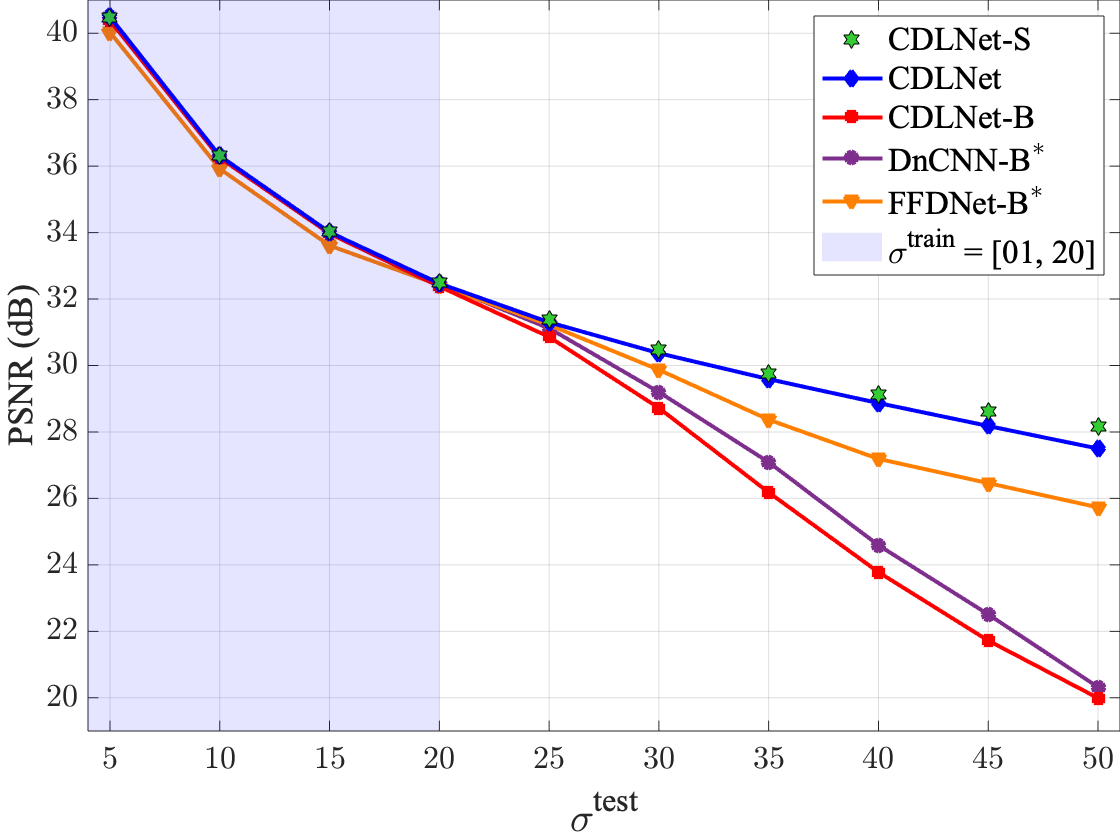}}
  \subfloat[PSNR plot $\sigma^{\mathrm{train}}={[20,30]}$ \label{fig:ColorBlindPlot2030}]{%
        \includegraphics[width=0.49\linewidth]{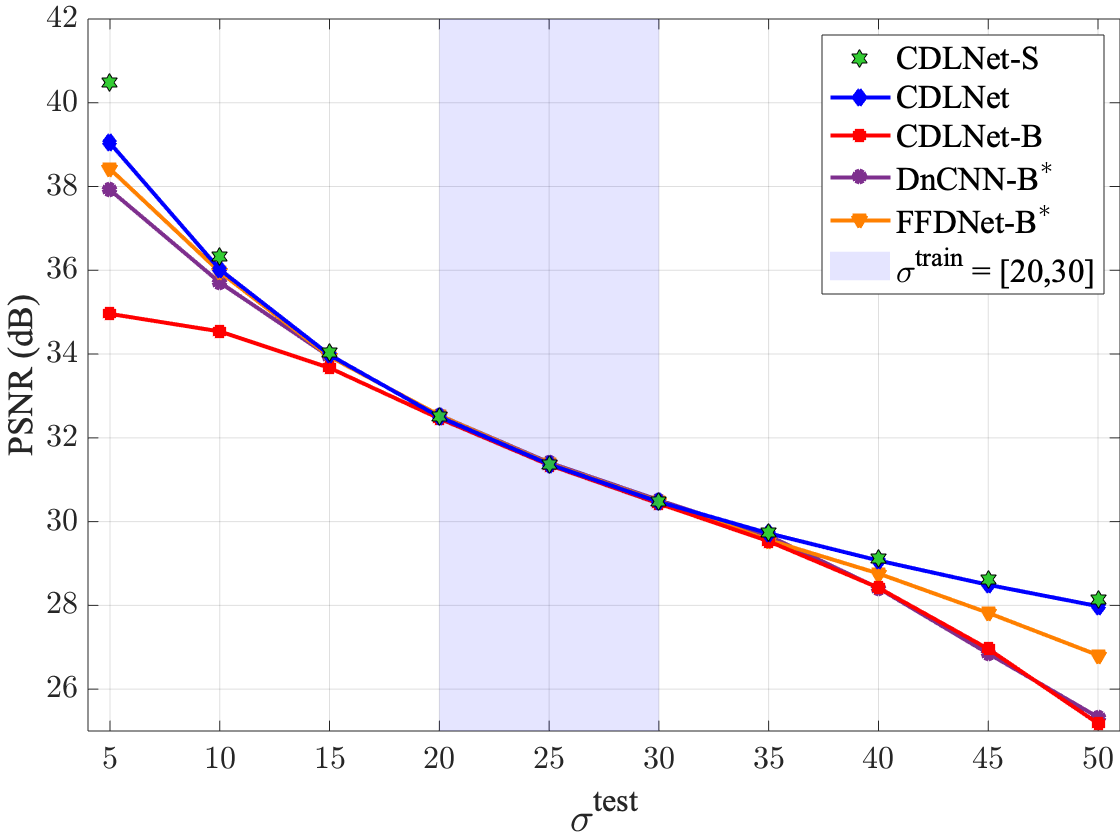}}
    \\\vspace*{8pt}
     \includegraphics[width=\linewidth]{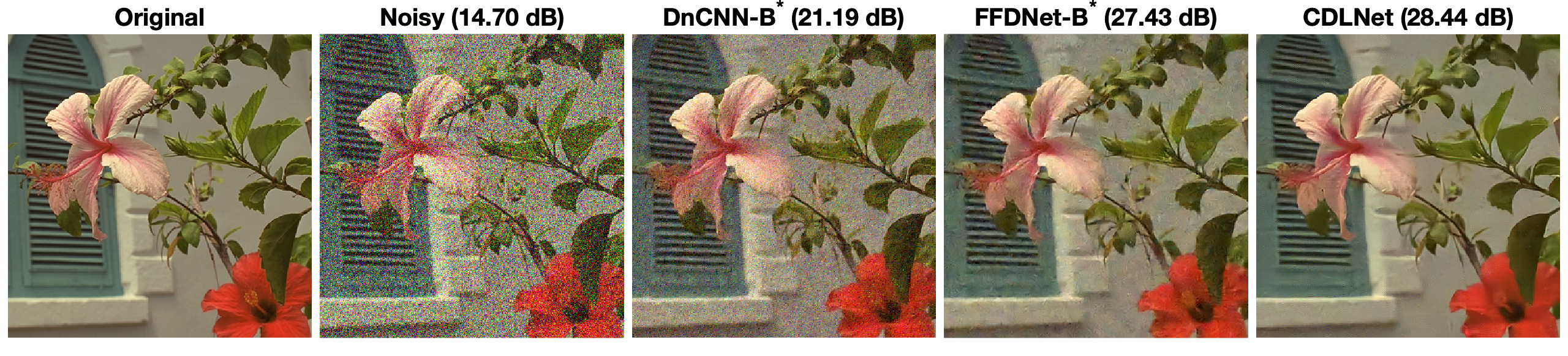}
    \\\vspace*{-10pt}
    \subfloat[Visual comparison for $\sigma^{\mathrm{train}}={[01,20]}$ and $\sigma^{\mathrm{test}}={50}$ \label{fig:ImgColor0120}]{%
       \includegraphics[width=\linewidth]{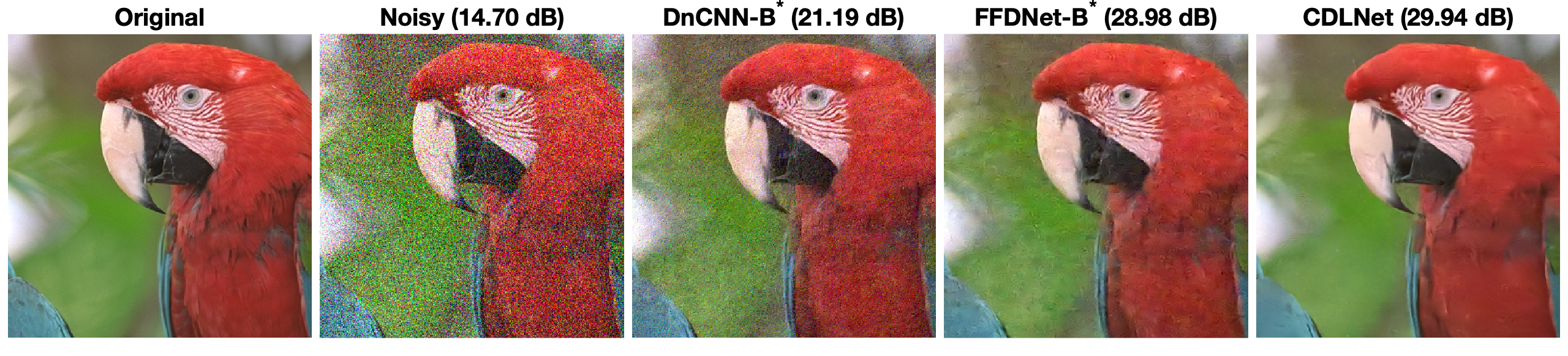}}
    \\\vspace*{8pt}
    \includegraphics[width=\linewidth]{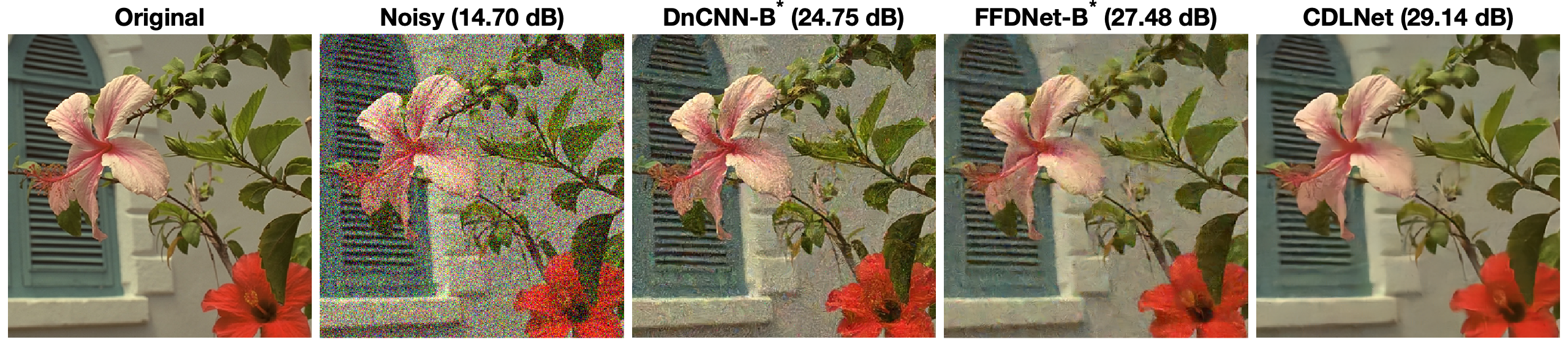}
    \\\vspace*{-10pt}
    \subfloat[Visual comparison for $\sigma^{\mathrm{train}}={[20,30]}$ and $\sigma^{\mathrm{test}}={50}$ \label{fig:ImgColor2030} ]{%
       \includegraphics[width=\linewidth]{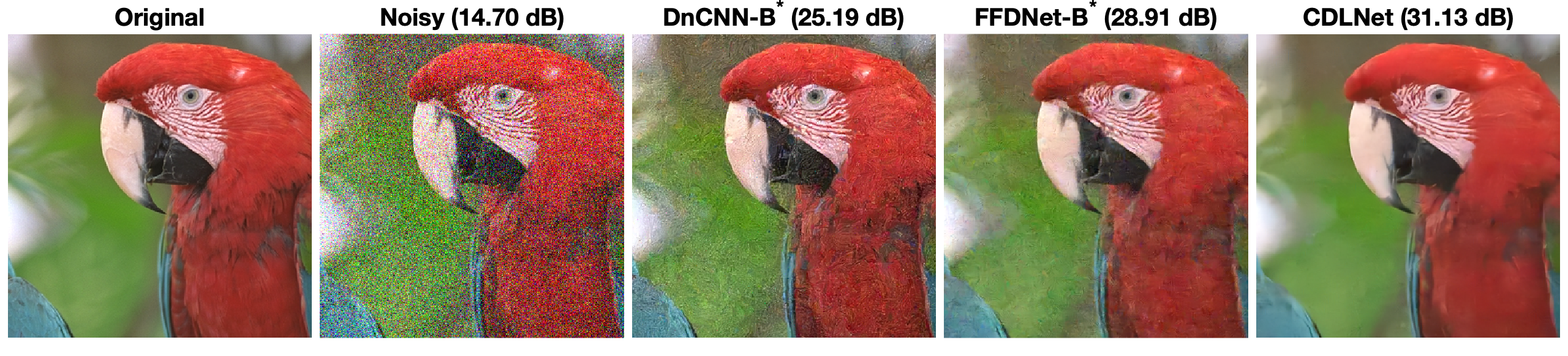}}
  \caption{(a,b) Performance of different color denoising networks trained on $\sigma^{\mathrm{train}}$ and tested on different $\sigma^{\mathrm{test}}$. Average PSNR calculated over BSD68 \cite{bsd}. (c,d) Visual comparison of different networks tested on noise-level $\sigma^{\mathrm{test}}=50$. Details are better visible by zooming.}
  \label{fig:ImgColor} 
\end{figure}

In this section, we consider the blind denoising and generalization scenarios, and compare the models equipped with the proposed adaptive threshold scheme to other models. In Figures \ref{fig:GrayBlindPlot0120} and \ref{fig:ColorBlindPlot0120}, we show the performance of the models trained on the noise range $\sigma^{\mathrm{train}}=[1,20]$ and tested on different noise-levels $\sigma^{\mathrm{test}}\in[5,50]$ for grayscale and color images, respectively. We trained the blind denoising version of DnCNN \cite{DnCNN}, DnCNN-B, and FFDNet on $\sigma^{\mathrm{train}}=[01,20]$, denoted as DnCNN-B$^\ast$ and FFDNet-B$^\ast$. The single points on these plots show the performance of CDLNet-S model with single noise-level training (i.e. $\sigma^{\mathrm{train}}=\sigma^{\mathrm{test}}$). Similarly, Figures \ref{fig:GrayBlindPlot2030} and \ref{fig:ColorBlindPlot2030}, show the performance results for the training noise-level $\sigma^{\mathrm{train}}=[20,30]$.

As shown in Figures \ref{fig:GrayBlindPlot0120} and \ref{fig:GrayBlindPlot2030}, all networks perform closely over the training noise-range. When tested on noise-levels outside the training range, the network with adaptive thresholds (CDLNet) significantly outperforms the other models. We observe that models without noise-adaptive thresholds have a very significant performance drop compared to the noise-adaptive model (CDLNet) when generalizing above the training noise-level, while CDLNet nearly matches the performance of the models trained for a specific noise-level (CDLNet-S) across the range. For Fig \ref{fig:GrayBlindPlot2030}, in spite of increasing input signal-to-noise ratio for noise-levels below the training range, we  observe that models without noise-adaptive thresholds have diminishing performance returns (note the plateau of CDLNet-B and DnCNN-B$^{\ast}$ in $\sigma^{\mathrm{test}}=[5,20]$). Even FFDNet-B$^\ast$, which takes the noise-level as input, fails to generalize, although with a smaller performance drop. On the other hand, the generalization behavior of CDLNet extends to the lower noise-range. At $\sigma^{\mathrm{test}}=5$, we notice a reduced performance of CDLNet compared to CDLNet-S. This may be explained by the need for a different thresholding model when the signal variance is much greater than that of the noise \cite{Mallat}.

Figures \ref{fig:GrayBlindPlot0120Small} and \ref{fig:GrayBlindPlot2030Small} show the noise-level generalization behavior of CSCNet\cite{Simon2019} and small CDLNet networks with and without adaptive thresholds. We trained the blind denoising version of CSCNet on $\sigma^\mathrm{train}=[01,20]$ and $\sigma^\mathrm{train}=[20,30]$, denoted CSCNet-B$^\ast$. We observe that small-CDLNet is also able to generalize like its larger paramter counter-part (CDLNet). Both CSCNet-B$^\ast$ and small-CDLNet-B are unable to generalize outside of their training noise-level range, due to the lack of noise-adaptivity in their architectures.

As shown in Figures \ref{fig:ColorBlindPlot0120} and \ref{fig:ColorBlindPlot2030} for the color denoising case, all networks perform closely over the training noise-range. Above the training range, we observe similar behavior as the in grayscale case: CDLNet generalizes whereas the other networks fail (DnCNN-B$^\ast$, CDLNet-B) or significantly drop in performance (FFDNet-B$^*$). However, below the training noise-level range (Fig. \ref{fig:ColorBlindPlot2030}), we note that FFDNet-B$^*$ and DnCNN-B$^\ast$ are able to generalize better than CDLNet-B, though still not as well as CDLNet.  

Visual comparisons of the denoising generalization of CDLNet, DnCNN-B$^\ast$, and FFDNet-B$^\ast$ are shown in Figures \ref{fig:ImgGray}, \ref{fig:ImgColor}. Our proposed adaptive model provides visually appealing results at the unseen noise-level ($\sigma^{\mathrm{test}}=50$), while DnCNN-B$^\ast$ fails to generalize and FFDNet-B$^\ast$ produces unwanted artifacts.

We further compare the blind denoising and generalization capabilities of the proposed method in Table \ref{table:generalization}. As observed in \cite{Mohan2020}, the BFCNN model has reduced performance in the training range compared to DnCNN-B while avoiding the failure outside the training range. CDLNet outperforms DnCNN-B inside the training range and also provides improved generalization outside the range compared to BFCNN \cite{Mohan2020}. Note that CDLNet has minor performance drop compared to a single noise-level mode (CDLNet-S) but this reduction is less significant than that of BFCNN \cite{Mohan2020}. CDLNet's performance distinguishes itself from that of FFDNet-B$^\ast$ in its superior generalization above the training noise-level range ($\sigma = 75$ in Table \ref{table:generalization}).

\begin{table}[ht]
\centering
\caption{Blind denoising and generalization comparison for grayscale images. All models are trained on $\sigma^{\mathrm{train}}=[0,55]$
. PSNR values averaged over BSD68 dataset \cite{bsd}.}
\resizebox{0.85\linewidth}{!}{%
\begin{tabular}{cccccc}\hline
\multirow{2}{*}{Model} & \multicolumn{5}{c}{Noise-level ($\sigma$)} \\
 & 5 & 15 & 25 & 50 & 75 \\ \hline
DnCNN-B \cite{DnCNN}     & 37.65 & \underline{31.60} & \underline{29.15} & 26.22 & 18.74  \\
BFCNN \cite{Mohan2020}   & \underline{37.72} & 31.58 & 29.12 & 26.17 & \underline{24.63}  \\
FFDNet-B$^\ast$ & 37.59 &  31.57 & 29.13 & \underline{26.27} & 23.76 \\
CDLNet & \textbf{37.75} & \textbf{31.63} & \textbf{29.20} & \textbf{26.31} &\textbf{24.80} \\ \hline
\end{tabular}%
}
\label{table:generalization}
\end{table}

\textbf{Effect of Noise-level Estimation Algorithms}:
Table \ref{table:noise_est} shows the difference in denoising performance between using the ground-truth (GT) noise-level and the two aforementioned noise-level estimation algorithms. The PCA based algorithm \cite{Liu2013} allows us to attain near ground-truth denoising performance at the cost of increased computation. The wavelet based estimation method, MAD \cite{Chang2000}, offers essentially no computational overhead, but significantly decreases denoising performance at the lower noise-level ranges.

\begin{table}[ht]
\centering
\caption{Effect of noise estimation algorithm on performance of CDLNet. PSNR values and inference timing averaged over the BSD68 dataset \cite{bsd}.}
\resizebox{\linewidth}{!}{%
\begin{tabular}{ccccclc}\hline
\multirow{2}{*}{Est. Algo.} & \multicolumn{5}{c}{Noise-level ($\sigma$)} & \multirow{2}{*}{GPU time} \\
 & 5 & 15 & 25 & 50 & 75 &  \\ \hline
GT & 37.75 & 31.63 & 29.20 & 26.31 & 24.80 & 13 ms \\
PCA \cite{Liu2013}& 37.73 & 31.62 & 29.20 & 26.30 & 24.76 & 23 ms \\
MAD \cite{Chang2000}& 37.18 & 31.55 & 29.18 & 26.30 & 24.75 & 13 ms \\\hline
\end{tabular}%
}
\label{table:noise_est}
\end{table}

\subsection{Joint Denoising and Demosaicing}
\begin{figure}[hb] 
    \centering
    \subfloat[PSNR plot $\sigma^{\mathrm{train}}={[01,20]}$ \label{fig:dmsc_generalize}]{%
        \includegraphics[width=0.55\linewidth]{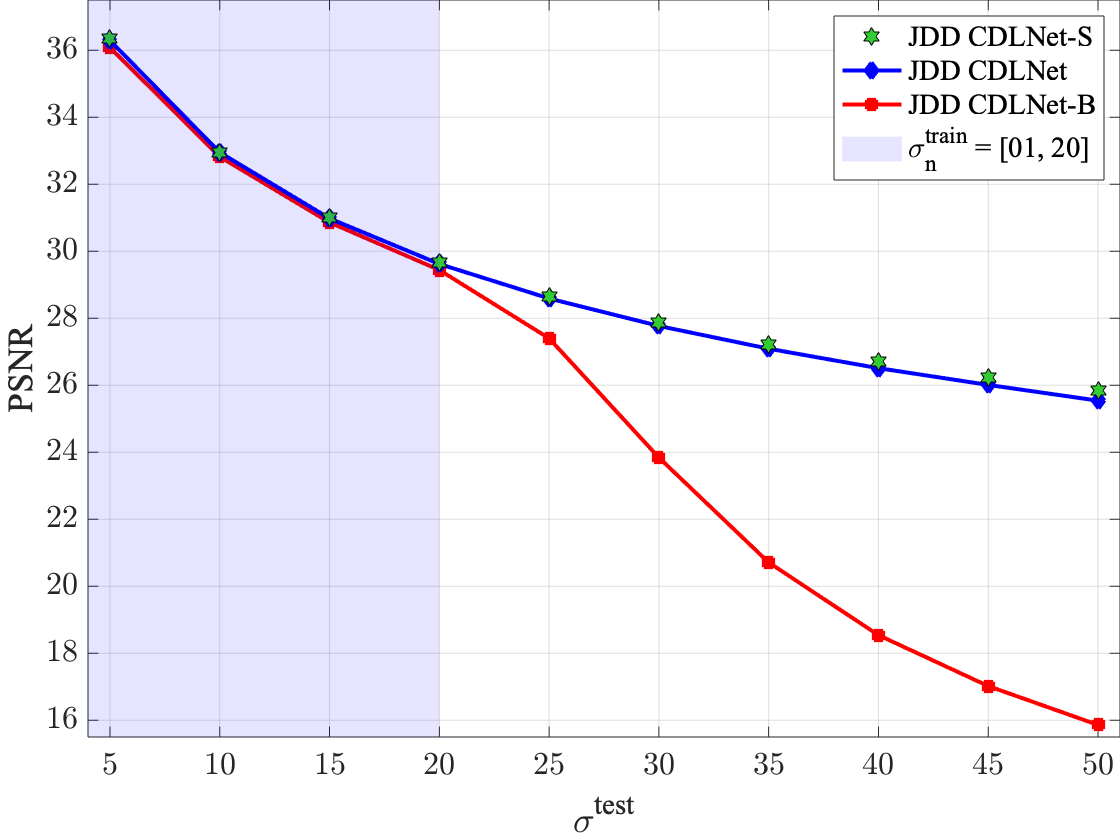}}
    \\\vspace*{2pt}
     \subfloat[Visual comparison\label{fig:DMSC0120Img09} ]{%
      \includegraphics[width=\linewidth]{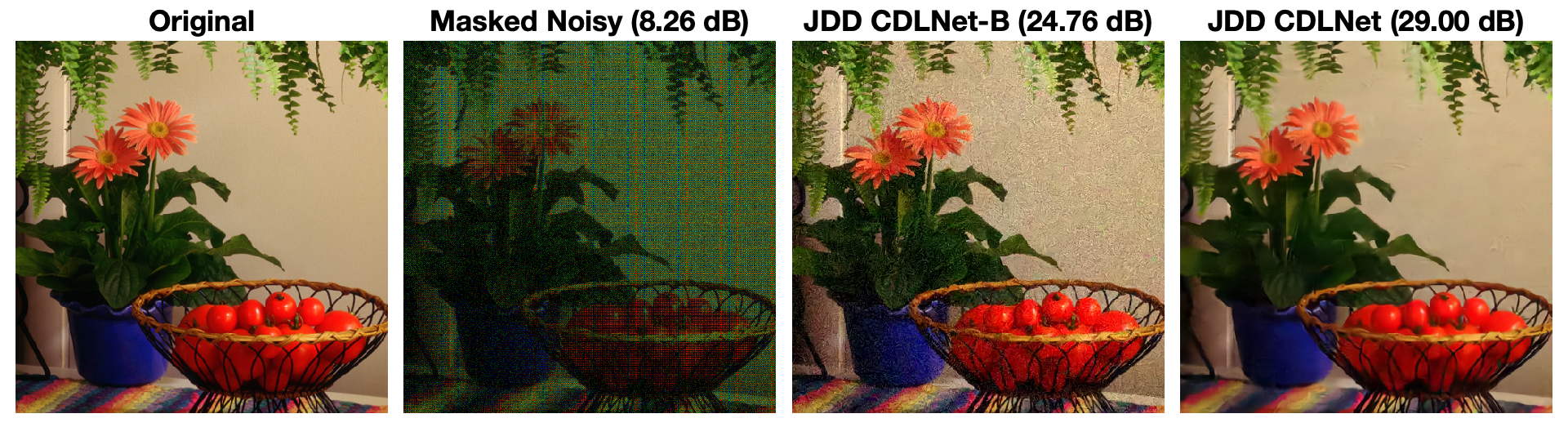}}
  \caption{(a) Performance of demosaicing CDLNets trained on $\sigma^{\mathrm{train}}=[01,20]$ and tested on different $\sigma^{\mathrm{test}}$. CDLNet and CDLNet-B are models with and without adaptive thresholds respectively. Average PSNR calculated over CBSD68 \cite{bsd}. (b) Visual comparison of JDD CDLNet and JDD CDLNet-B tested on noise-level $\sigma^{\mathrm{test}}=30$. Details are better visible by zooming.}
  \label{fig:ImgDMSC} 
\end{figure}

In this section, we demonstrate that the performance and generalization characteristics of CDLNet extend beyond image denoising and into the realm of linear inverse problems. Specifically, we look at the task of jointly denoising and demosaicing (JDD) color images acquired with a Bayer mask. First, we verify the efficacy of the proposed adaptive thresholds. Figure \ref{fig:dmsc_generalize} and \ref{fig:DMSC0120Img09} show the generalization performance of CDLNet on the JDD task across noise-levels, confirming the characteristic behavior of the adaptive thresholds observed in the grayscale and color denoising scenarios extends to the JDD case. 

\begin{figure}[ht]
\centering
\includegraphics[width=\linewidth]{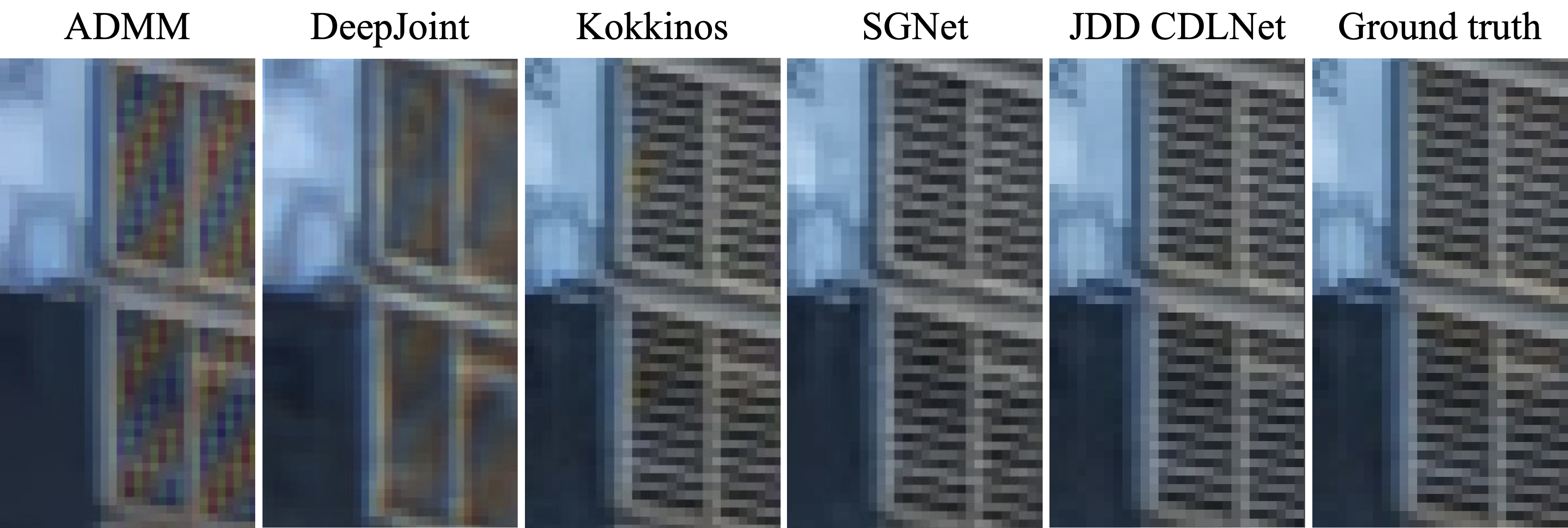}
\caption{Visual comparison of JDD CDLNet and \soa joint denoising and demosaicing methods on a crop of image 270 from the MIT moir\'{e} dataset \cite{Gharbi2016} with $\sigma=0$. Results from other methods are obtained from \cite{Liu_2020_CVPR}.}
\label{fig:DmscCompare}
\end{figure}

Table \ref{tab:dmsc} compares the JDD performance of our proposed CDLNet model (trained on $\sigma\in[01,20]$) in low and high parameter count regimes (JDD small-CDLNet and JDD CDLNet, resp.), versus competing DNNs (SGNet\cite{Liu_2020_CVPR}, Kokkinos\cite{Kokkinos_2018_ECCV}, DeepJoint\cite{Gharbi2016}) and the classical method ADMM \cite{Tan2017} baseline across a range of noise-levels. As with the proposed JDD CDLNet models, each deep-learning based method has trained a single model to perform across the range of presented test noise-levels ($\sigma\in\{5,10,15\}$). These compared deep-learning methods \cite{Liu_2020_CVPR,Kokkinos_2018_ECCV,Gharbi2016} each follow the FFDNet \cite{FFDNet} approach of noise-level adaptivity by presenting their network with the input noise-level $\sigma$ concatenated with the input. We observe that CDLNet outperforms all methods when scaled close to the parameter count of SGNet \cite{Liu_2020_CVPR}, and even outperforms other methods when scaled to the lower parameter count range (JDD small-CDLNet). A visual comparison of these methods on a tough patch from the MIT moir\'{e} dataset \cite{Gharbi2016} is given in Figure \ref{fig:DmscCompare}. Again, we see CDLNet on par with the \soa SGNet \cite{Liu_2020_CVPR}, where other methods introduce color and texture artifacts.

\begin{table}[ht]
	\caption{
	Comparison of CDLNet models in low and high parameter count regime against \soa methods on the JDD task. Reconstruction PSNR over the Urban100\cite{Urban100}/MIT moir\'{e}\cite{Gharbi2016} datasets is shown. PSNRs reported in \cite{Liu_2020_CVPR}. 
	}
	\begin{center}
    \resizebox{0.99\linewidth}{!}{%
	\begin{tabular}{ccccccc}
	\hline
	\multirow{2}{*}{Model} & \multirow{2}{*}{Params} & \multicolumn{3}{c}{Noise-level ($\sigma$)} \\
	 & & 5 & 10 & 15 \\ \hline
	ADMM~\cite{Tan2017} & - & 28.57 / 28.58 & 27.48 / 28.26 & 28.37 / 27.58 \\
	Kokkinos~\cite{Kokkinos_2018_ECCV} & 380k & 34.07 / 31.94 & 31.73 / 30.01 & 29.87 / 28.28 \\
	DeepJoint~\cite{Gharbi2016} & 560k & 34.04 / 31.82 & 31.60 / 29.75 & 29.73 / 28.22 \\
	SGNet\cite{Liu_2020_CVPR} & 795k & \underline{34.54} / \underline{32.15} & 32.14 / 30.09 & 30.37 / 28.60 \\
	JDD small-CDLNet & 373k & 34.43 / 32.05 & \underline{32.23} / \underline{30.16} & \underline{30.68} / \underline{28.81} \\
	JDD CDLNet & 796k & {\bf 34.60} / {\bf 32.16} & {\bf 32.42} / {\bf 30.27} & {\bf 30.89} / {\bf 28.94} \\ \hline
	\end{tabular}%
	}
	\end{center}

	\label{tab:dmsc}
\end{table}

\subsection{Unsupervised learning using MC-SURE Loss}
\begin{figure}[ht]
    \centering
    \subfloat[PSNR plot $\sigma^{\mathrm{train}}={[01,20]}$ \label{fig:GrayMCSURE0120}]{%
        \includegraphics[width=0.49\linewidth]{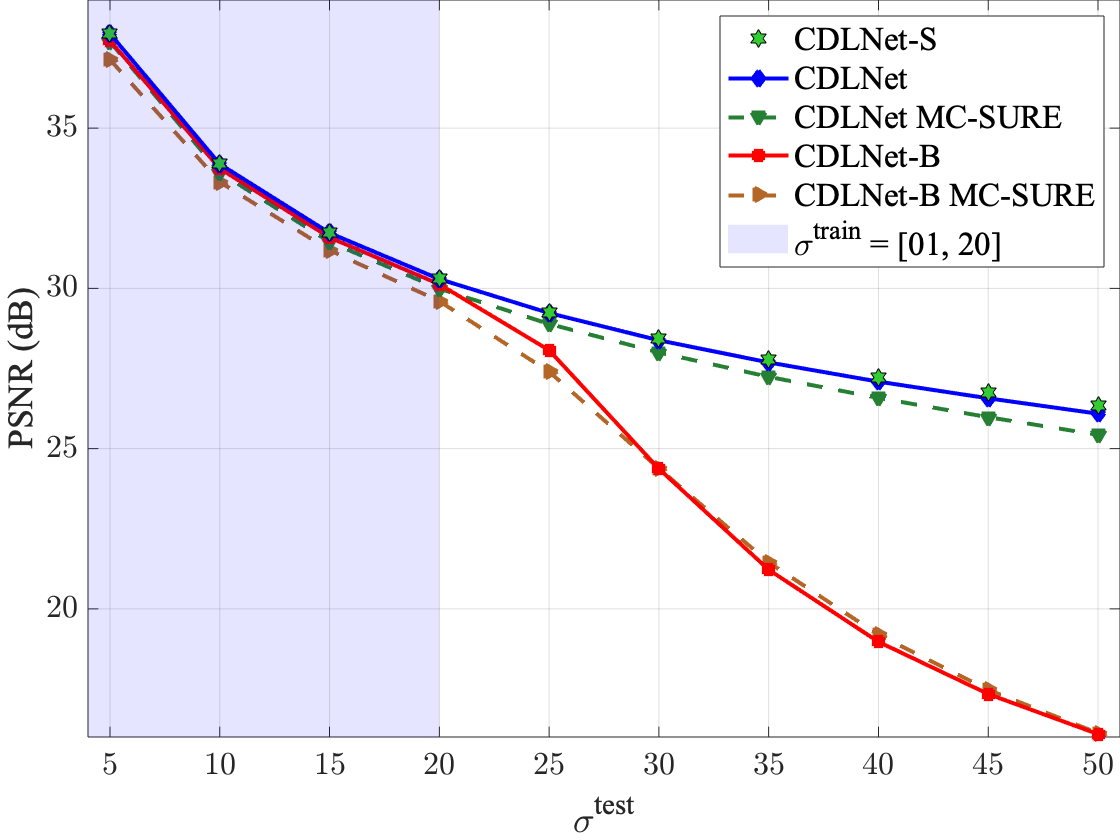}}
  \subfloat[PSNR plot $\sigma^{\mathrm{train}}={[20,30]}$ \label{fig:GrayMCSURE2030}]{%
        \includegraphics[width=0.49\linewidth]{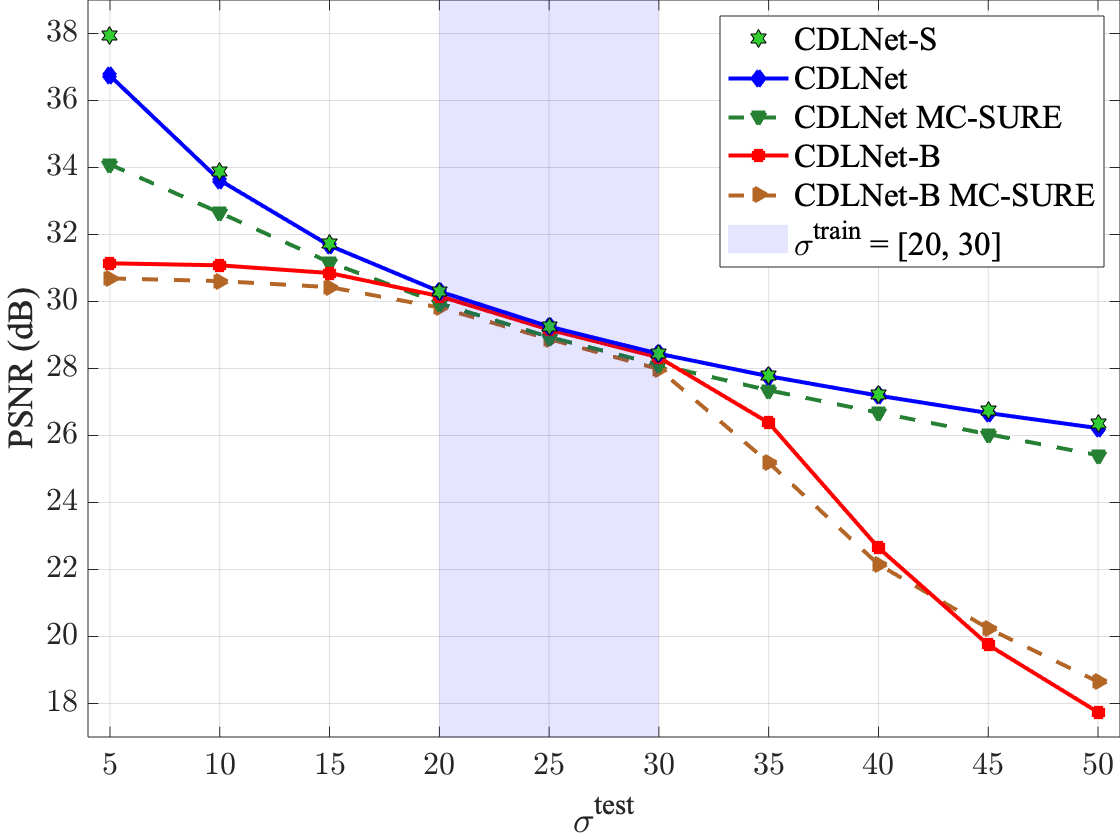}}
    \\\vspace*{8pt}
     \includegraphics[width=\linewidth]{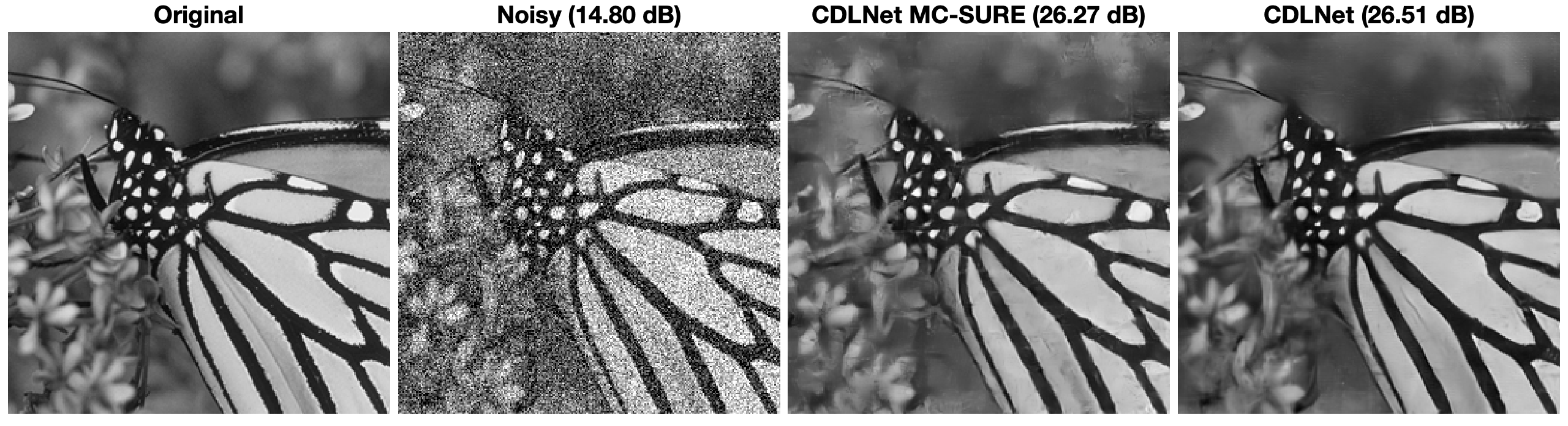}
    \\\vspace*{-10pt}
    \subfloat[Visual comparison for $\sigma^{\mathrm{train}}={[01,20]}$ and $\sigma^{\mathrm{test}}={50}$ \label{fig:ImgSURE0120}]{%
      \includegraphics[width=\linewidth]{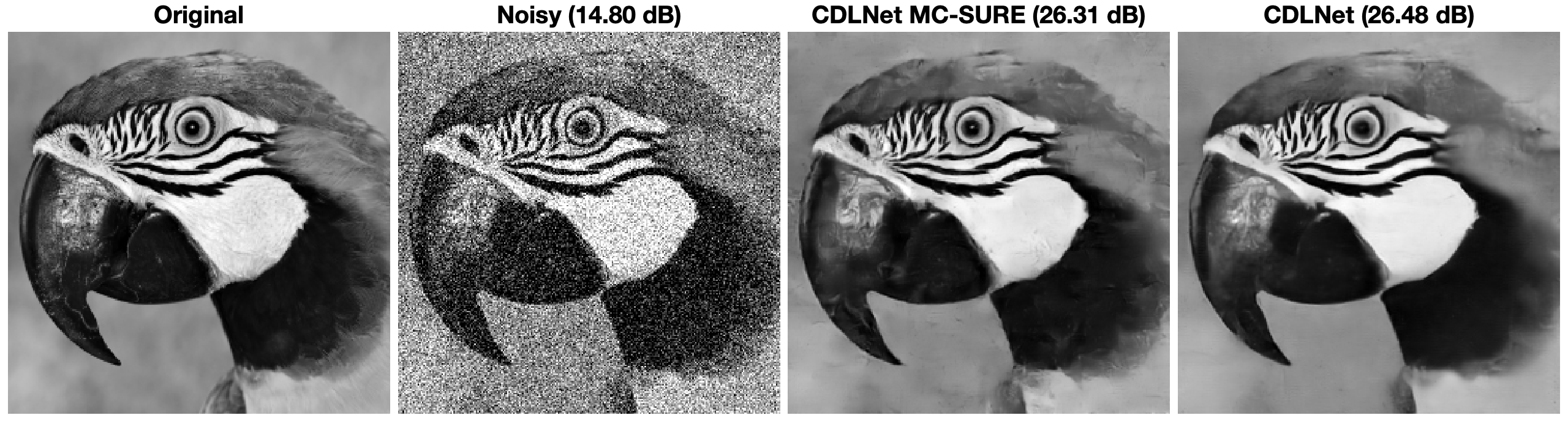}}
    \\\vspace*{8pt}
    \includegraphics[width=\linewidth]{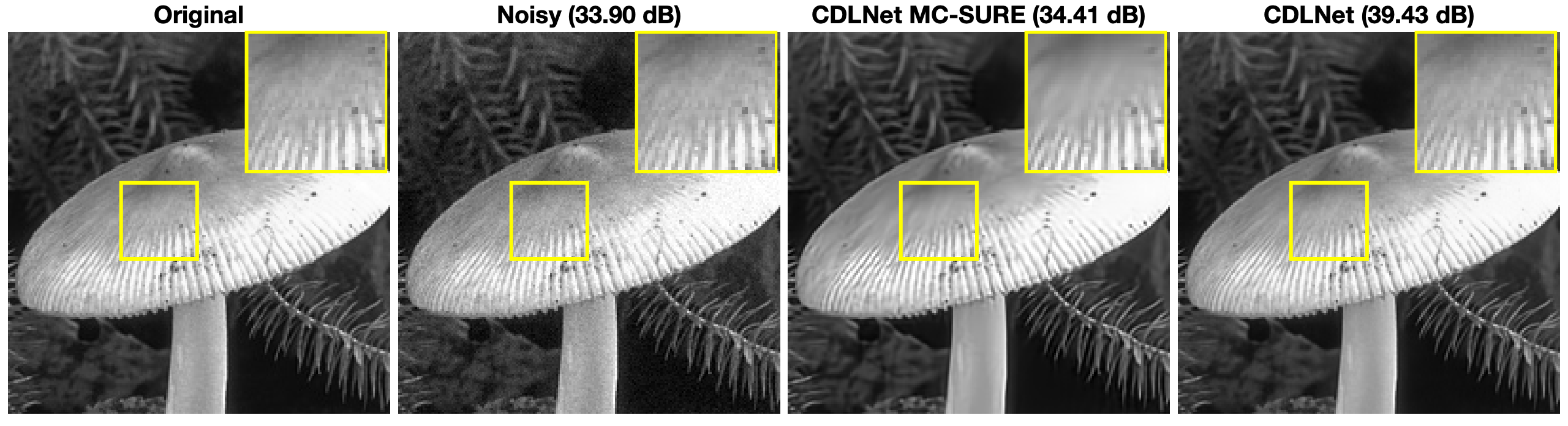}
    \\\vspace*{-10pt}
    \subfloat[Visual comparison for $\sigma^{\mathrm{train}}={[20,30]}$ and $\sigma^{\mathrm{test}}={5}$ \label{fig:ImgSURE2030} ]{%
      \includegraphics[width=\linewidth]{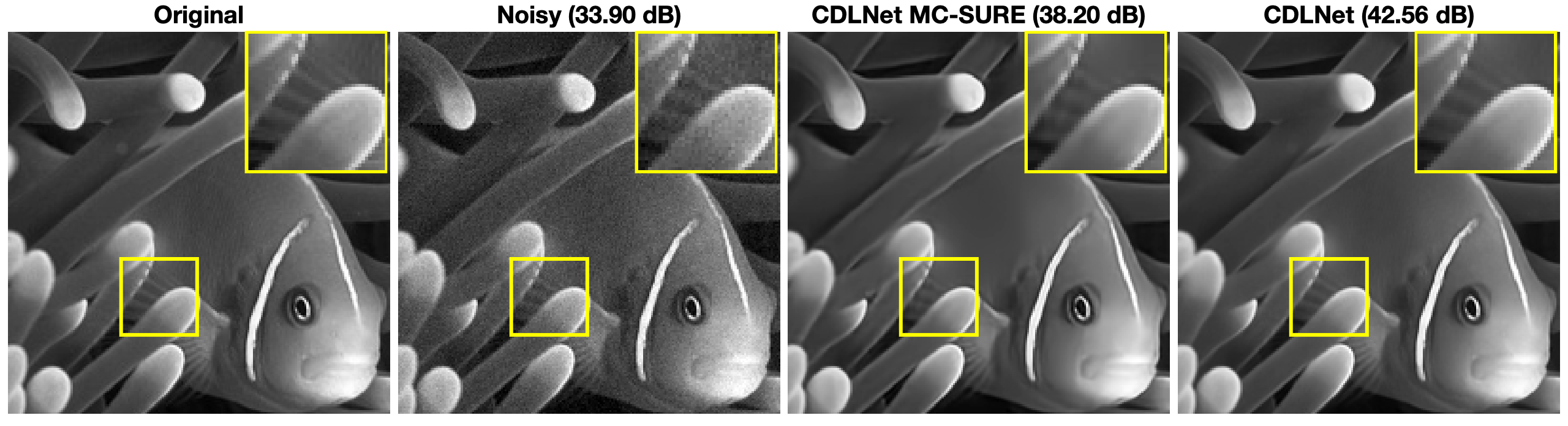}}
  \caption{(a,b) Performance of CDLNet trained on $\sigma^{\mathrm{train}}$ with MSE and MC-SURE loss functions and tested on different $\sigma^{\mathrm{test}}$. Average PSNR calculated over BSD68 \cite{bsd}. Visual comparison of different cases tested on noise-level (c) $\sigma^{\mathrm{test}}=50$ and (d) $\sigma^{\mathrm{test}}=5$. Details are better visible by zooming.}
  \label{fig:ImgMCSURE} 
\end{figure}

In this section, we compare the performance of CDLNet under supervised learning (using MSE loss) vs. unsupervised learning (using MC-SURE loss).
Figure \ref{fig:GrayMCSURE0120} shows the result of CDLNet and CDLNet-B when trained with MSE or MC-SURE loss functions with $\sigma^{\mathrm{train}} = [01,20]$ for grayscale image denoising. We observe that inside the training range, the networks trained with MC-SURE loss function perform slightly worse than the networks trained with MSE. This is expected since the MC-SURE loss function minimizes an unbiased estimate of the MSE, but does not require any ground-truth information. The networks with adaptive threshold scheme (CDLNet) show generalization across the noise-levels outside the training range for both cases of MSE and MC-SURE losses. 

In Figure \ref{fig:GrayMCSURE2030}, we show the result of CDLNet and CDLNet-B when trained with MSE or MC-SURE loss functions for $\sigma^{\mathrm{train}} = [20,30]$ with grayscale images. Generalization is observed in the adaptive MC-SURE trained model above the training range, however, a significant performance drop is observed below the training range (when compared to the MSE trained model). This change of behaviour is likely due the lack of texture/detail seen during unsupervised training in the noise-level range, whereas the MSE trained model is still able to update its weights on fine-detail and texture present in the ground-truth signal. Visual comparison of the results for $\sigma^{\mathrm{test}}=5$ is shown in \ref{fig:ImgSURE2030}. We note the undesired smoothing artifact on results for model trained with MC-SURE loss. Further evidence for this hypothesis is seen in the relative lack of texture filters in the dictionary of the unsupervised vs. supervised models, shown in Figure \ref{fig:AllGrayDictionary}(d, e).

\subsection{Comparison of learned dictionaries}

\begin{figure}[ht]
\centering
\includegraphics[width=0.95\linewidth]{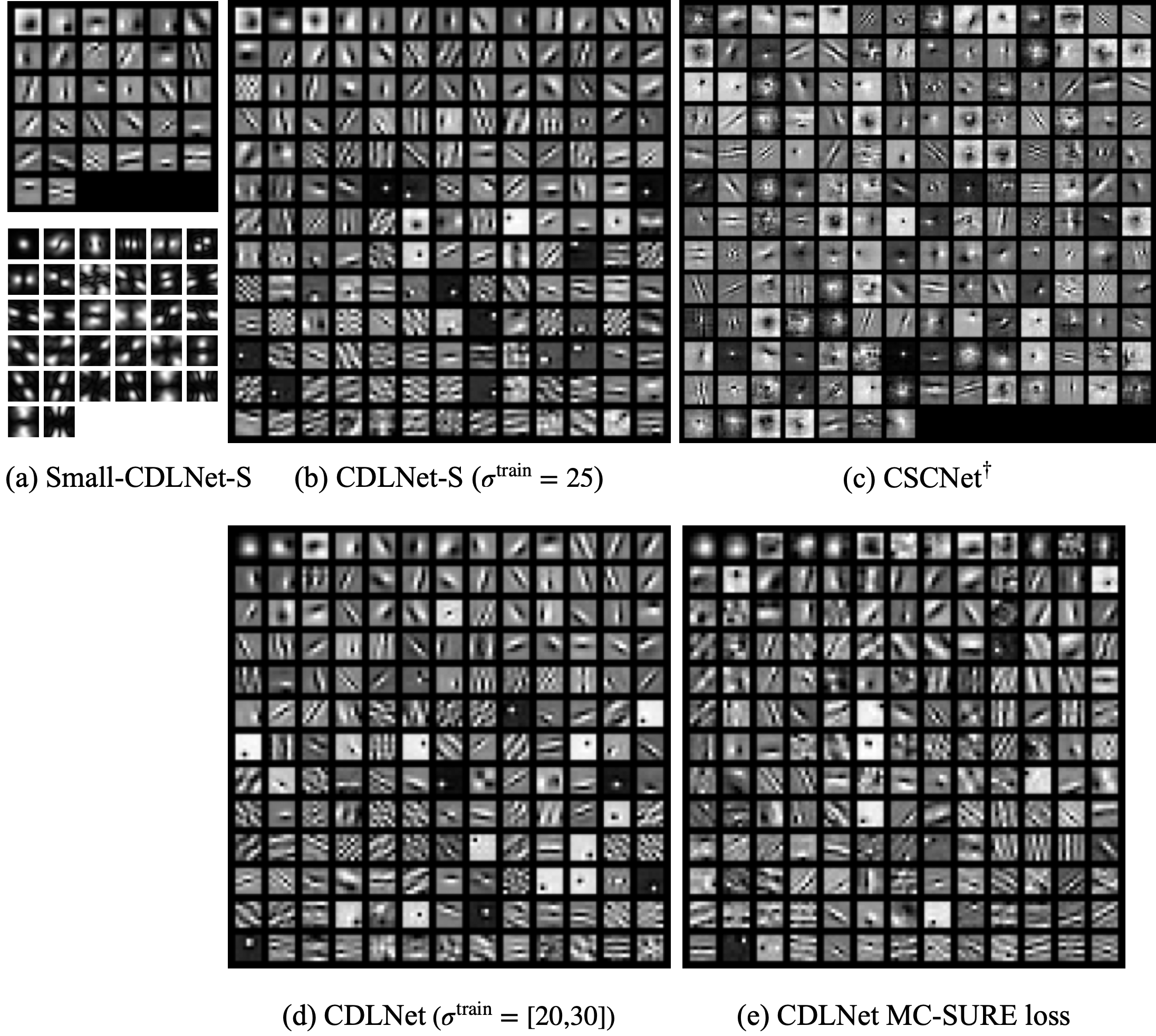}
\caption{Final learned dictionaries ($\bm{D}$) for (a) small-CDLNet-S (in spatial domain (top) and frequency domain (bottom)), 32 filters of size $7\times7$ (b) CDLNet-S, 169 filters of size $7\times 7$, and (c) CSCNet \cite{Simon2019}, 175 filters of size $11\times 11$ trained on $\sigma = 25$. $~\dagger$ CSCNet figure obtained from the models provided online by \cite{Simon2019}. (d) CDLNet filters trained on $\sigma^{\mathrm{train}}=[20,30]$ and (e) CDLNet filters trained with MC-SURE loss. CDLNet filters are ordered by relative usage over the BSD68 dataset \cite{bsd}.}
\label{fig:AllGrayDictionary}
\end{figure}

A side-by-side comparison of the final dictionaries ($\bm{D}$) for different
grayscale image denoising networks is shown in Figure
\ref{fig:AllGrayDictionary}. The convolutional dictionaries obtained from the CDLNet-S model and CSCNet, with the CDLNet models' filters ordered by average utilization over the BSD432 dataset \cite{bsd}. We see that CDLNet-S offers a greater diversity in its learned filters compared to CSCNet. The small-CDLNet-S model has mostly learned directional ``Gabor-like" filters, missing the finer granularity of directions present in the larger dictionary of the CDLNet-S model which can explain
the performance and parameter-count trade-off. 
We emphasize the similarity of the
learned filters for CDLNet-S, model trained on single noise-level
(\Fig\ref{fig:AllGrayDictionary}b), and CDLNet, model trained on an input
noise range (\Fig\ref{fig:AllGrayDictionary}d). Note that the
noise-adaptivity of CDLNet is explicitly modeled in the network
encoder and hence does not necessarily effect the learned dictionary
required for image reconstruction. Furthermore, we observe a relative lack of texture filters in the dictionary for CDLNet when trained with MC-SURE loss (\Fig\ref{fig:AllGrayDictionary}e) compared to the MSE loss (\Fig\ref{fig:AllGrayDictionary}d) which is presumably caused by absence of clean images during training.

\begin{figure}[ht]
\centering
\includegraphics[width=0.95\linewidth]{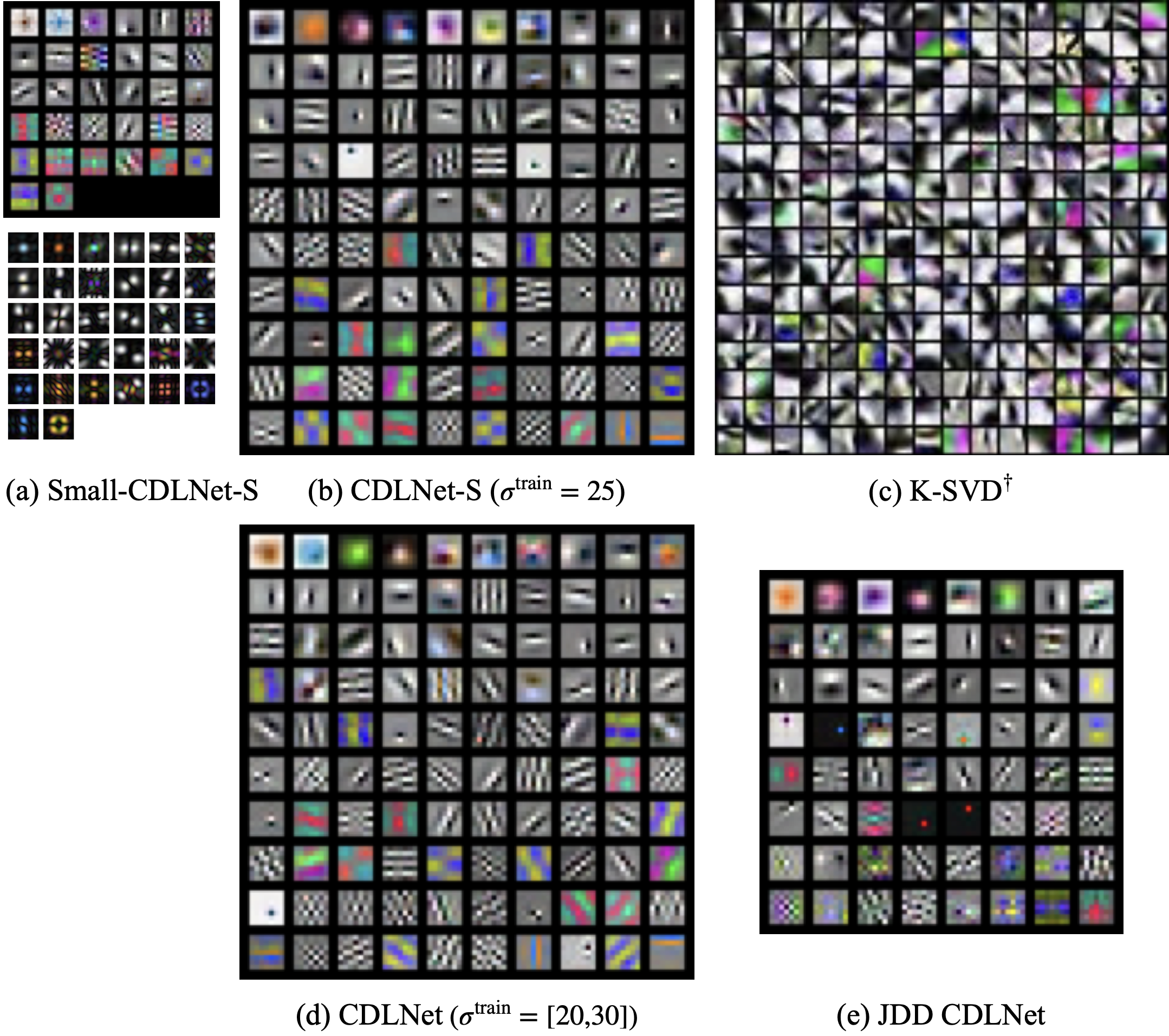}
\caption{Final learned dictionaries ($\bm{D}$) for color denoising (a) small-CDLNet-S (in spatial domain (top) and frequency domain (bottom)) (b) CDLNet-S, and (c) K-SVD \cite{rubinstein2008efficient}. $~\dagger$ K-SVD figure obtained from \cite{xu2015vector}. (d) CDLNet filters trained on $\sigma^{\mathrm{train}}=[20,30]$ and (e) JDD CDLNet filters. All CDLNet filters are ordered by relative usage over the CBSD68 dataset \cite{bsd}}
\label{fig:AllColorDictionary}
\end{figure}

Comparison of final dictionaries ($\bm{D}$) learned for different color denoising and JDD networks are shown in Figure \ref{fig:AllColorDictionary}. Similar to classical dictionary learning methods such as K-SVD \cite{rubinstein2008efficient}, CDLNet-S (\Fig\ref{fig:AllColorDictionary}a,b) learns a mixture of color and basic spatial structure filters. However, the structural elements are better presented in CDLNet-S dictionary compared to the classical method. As in the grayscale case, CDLNet-S (\Fig\ref{fig:AllColorDictionary}b) and CDLNet (\Fig\ref{fig:AllColorDictionary}d) have similar learned filters. Interestingly, colored edge filters are present in both small-CDLNet (\Fig \ref{fig:AllColorDictionary}a) and large parameter count CDLNet (\Fig\ref{fig:AllColorDictionary}b,d), although they are among the less utilized filters for CDLNet. In spite of their infrequent use, they are still vital for color image reconstruction and have a high impact on the MSE loss. Consequently, such filters are given precedence over fine details for the lower parameter count case. We also emphasize the mixture of color and basic structural filters in the case of JDD task (\Fig\ref{fig:AllColorDictionary}e).

\subsection{Discussion} \label{sec:discussion}
The interpretability of CDLNet is in part due to its derivation from a sparse-coding and dictionary learning framework. This allows us to understand the learned filters, intermediate representations, and parameters of the network. Other unrolled dictionary learning based networks such as\cite{Simon2019,Sreter2018,Lecouat2020Games,Scetbon2021}, share this property with CDLNet and stand in stark contrast to black-box DNNs (such as DnCNN \cite{DnCNN} and FFDNet \cite{FFDNet}). However, prior works based on unrolled dictionary learning consider learned parameter count as a measure of model complexity and focus on reducing this complexity while achieving performance close to black-box DNNs. For instance, in the case of natural image denoising, Simon et al. (CSCNet) \cite{Simon2019} and Lecouat et al. \cite{Lecouat2020Games} show significantly lower parameter count while having a performance gap compared to their \soa black-box DNNs. In Tables \ref{table:graysingle} and \ref{table:colorsingle}, we observe that our proposed method performs close to that of \soa fully convolutional neural networks with 3-8 times less parameters. However, unlike the previously mentioned networks, we also show that our proposed method can surpass the denoising performance of \soa FCNNs (DnCNN \cite{DnCNN}, FFDNet \cite{FFDNet}) when scaled to a similar parameter count. These results suggest that the performance gap observed in other methods may be due to their relatively low parameter count, however, unlike CDLNet they lack a computationally manageable framework to realize this scaling of network size.
We emphasize that although the scalable performance of CDLNet is afforded by small-strided convolutions and untying the weights between unrolling iterations, the interpretation of the network as an approximate sparse-coding module is no less valid than with tied weights, such as in LKSVD \cite{Scetbon2021}, CSCNet \cite{Simon2019}, and ACSC \cite{Sreter2018}, and each case falls under the theoretical framework of Chen et al. \cite{Chen2018}.

Beyond network complexity, which has been the primary focus in the literature, we show interpretable network construction leads to noise-adaptivity within CDLNet. Since the CDLNet architecture is derived from a sparse-coding framework, the thresholds can be related to the observation noise-level. We leverage this relation in our design to explicitly model how the operations of the network adapt to the input noise-level within each layer. Figures \ref{fig:ImgGray} and \ref{fig:ImgColor} show that this explicit noise adaptive model allows for a near-perfect generalization of the network's performance to unseen noise-levels at inference, whereas the implicitly defined noise-adaptivity of black-box DNNs such as DnCNN \cite{DnCNN} and FFDNet \cite{FFDNet} fail in comparison. We further show that this noise generalization characteristic is not exclusive to purely denoising, and  similar results may be obtained when CDLNet is adapted to unsupervised learning (via the MCSURE loss function) and joint denoising and demosaicing (via incorporating the observation operator).

\section{Conclusion} \label{sec:conclusion}
In this work we presented CDLNet, a DNN for natural image denoising derived from convolutional dictionary learning and constructed without use of tricks from the standard deep learning tool-box. 
We showed performance close to \soa fully convolutional neural networks in natural image denoising at a relatively low learned parameter count. By using small-strided convolutions and increasing the number of unrollings (depth) and filters (width), we further showed that CDLNet can outperform \soa networks when scaled to a similar parameter count.
Similarly, in the joint denoising and demosaicing task, CDLNet achieves \soa performance.
Additionally, we leveraged the interpretable network construction to explicitly model the relation between the thresholds in each layer and the input noise-level. We observed near perfect generalization on noise-levels outside the training range for denoising, joint denoising and demosaicing, and unsupervised learning. This shows an example of how interpretable construction can be used to further improve network design. 

In future work, we aim to consider noise-adaptive generalization in other inverse problems such as compressed sensing. Additionally, the proposed framework can be applied to imaging modalities other than natural images. Specifically, the unsupervised training and generalization can be helpful in medical image denoising tasks where ground-truth samples may be unavailable. 
As a broader impact, further use of signal processing models within deep neural network architectures to yield novel capabilities is an exciting avenue of future research.

\section*{Appendix}
\subsection{Alternate formulation of color denoising via block-thresholding} \label{sec:app_bt}
\begin{figure}[ht]
\centering
\includegraphics[width=0.95\linewidth]{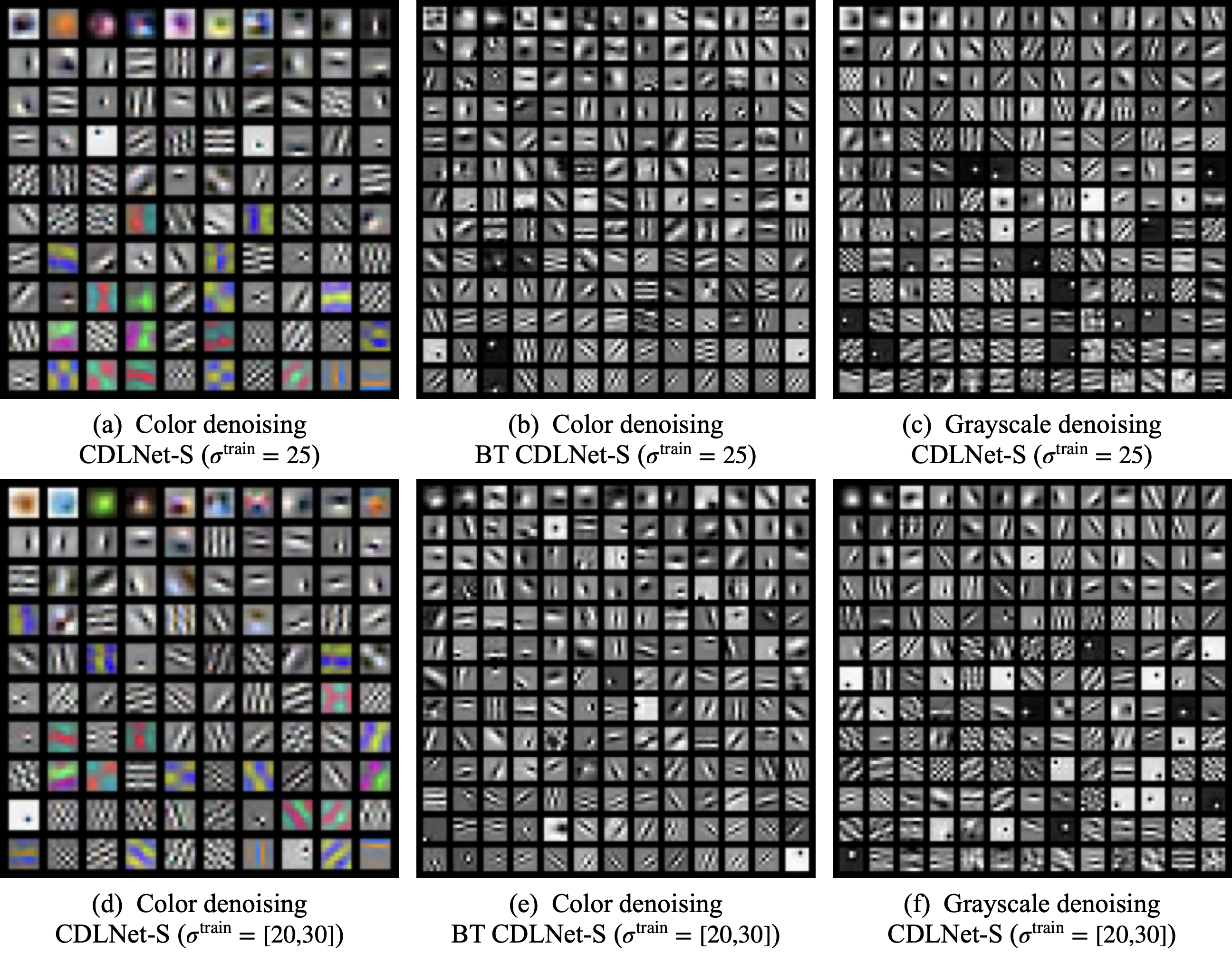}
\caption{Final learned dictionaries ($\bm{D}$) for denoising tasks. Filters are ordered by their relative usage over the (C)BSD68 dataset \cite{bsd}.}
\label{fig:BTDictionary}
\end{figure}

In the above presentation of CDLNet, we chose to represent color-images via a color dictionary acting on a grayscale subband images. 
Here we consider an alternate formulation in which a single grayscale dictionary models the structural information of each RGB channel individually. Each color component of the reconstructed image is represented by its own set of $M$ subband images convolved with the synthesis dictionary, and we refer to this subband representation as having ``color coefficients", $\z^m[n] \in \R^3, ~ \forall m=1,2,\dots,M$. The intuition for this choice of dictionary and representation can be seen in the fact the most of the learned filters of the color denoising CDLNet (Figure \ref{fig:BTDictionary}) turn out to be grayscale. Going back to the BPDN formulation \eqref{eqn:bpdn}, it no longer makes sense to seek $\ell_1$ sparsity of the entire subband representation as we do not want to encourage the color coefficients to be sparse, i.e. we want the color coefficients to use any mixture of R, G, B components without penalty. This property may be encoded by the group norm regularizer, $\norm{\z}_{2,1} \coloneqq \sum_{n,m} \norm{\z^m[n]}_2$. Our BPDN formulation then changes accordingly to,
\begin{equation} \label{eqn:bpdn_bt}
\underset{\z}{\mathrm{minimize}} ~ \frac{1}{2}\norm{\y -\D\z}_2^2 + \lambda \norm{\z}_{2,1}.
\end{equation}
Subsequently, the soft-thresholding in ISTA is replaced with block-thresholding,
\begin{equation} \label{eqn:ibta}
\z^{(k+1)} \coloneqq \mathrm{BT}(\z^{(k)} - \eta \D^\top(\D\z^{(k)} - \y), \,
\eta\lambda),
\end{equation}
where 
$${\mathrm{BT}(\x,\tau)^m[n] \coloneqq \frac{1}{\norm{\x^m[n]}_2}\x^m[n]\max(0, \norm{\x^m[n]}_2 - \tau)}$$
is the subband and pixel-wise block-thresholding operator \cite{khalilian2019strip,ng2010solving}. 

\textbf{Unrolling}: Unrolling the above block-thresholding formulation of convolutional dictionary learning for color image denoising involves using learned convolution analysis and synthesis filterbanks with $M$ subbands, and applying each filterbank identically on the RGB components of the images (and latent representation). Consequently, a block-thresholding CDLNet (BT CDLNet) of equivalent unrollings, filters, and stride to a soft-thresholding network will have the same computational complexity but $3$ times the memory requirement for storing the subband images. Note that this is not equivalent to simply processing each channel of the RGB input image separately, as the block-thresholding in each layer jointly processes the RGB coefficients.

Noise adaptivity can be proposed in exactly the same way, with learned thresholds $\boldsymbol{\tau}^{(k)} \in \R^M$ following the previously proposed affine noise-level model \eqref{eqn:thresh}.

\textbf{Experiments}: The following section uses the same experimental setup and naming convention as Section \ref{sec:results}, with the addition of block-thresholding networks denoted by the prefix ``BT". Block-thresholding CDLNet variants were trained in both low and high parameter count regimes, with parameters $(K,M,\sqrt{P},s) = (30,64,7,1),~(40,169,7,2)$, respectively.

Table \ref{table:colorsingle_BT} shows the results of the block-thresholding models trained for single noise-levels added to Table \ref{table:graysingle}. We see that the block-thresholding network (BT small-CDLNet-S) is on-par with the soft-thresholding (small CDLNet-S) in the small parameter count regime. In the large parameter count regime, we see the block-thresholding network (BT CDLNet-S) has a consistent performance gain over CDLNet-S. This may be due to the slightly larger number of the parameters used in BT CDLNet over CDLNet, or the redistribution of parameters between filters and layers.

\begin{table}[ht]
\centering
\caption{Color image denoising performance (PSNR) on CBSD68 testset ($\sigma = \sigma^{\mathrm{train}} = \sigma^{\mathrm{test}}$). All learned models trained on CBSD432\cite{bsd}, except CSCNet (CBSD432\cite{bsd} $+$ Waterloo ED \cite{ma2017waterloo}). PSNRs reported in respective citations.}

\resizebox{\linewidth}{!}{%
\begin{tabular}{cccccccc} \hline
\multirow{2}{*}{Model} & \multirow{2}{*}{Params} & \multicolumn{4}{c}{Noise level ($\sigma$)} \\
 & & 5 & 15 & 25 & 50 \\ \hline
CBM3D  &  -                    & 40.24 & 33.49 & 30.68 & 27.36 \\
CSCNet \cite{Simon2019} & 186k & -     & 33.83 & 31.18 & 28.00 \\
small-CDLNet-S          & 190k & 40.46 & 33.96 & 31.30 & \underline{28.05} \\
BT small-CDLNet-S       & 190k & 40.45 & 33.98 & 31.28 & 28.03 \\
FFDNet \cite{FFDNet}    & 486k & -     & 33.87 & 31.21 & 27.96 \\
DnCNN \cite{DnCNN}      & 668k & \bf{40.50} & \underline{33.99} & \underline{31.31} & 28.01 \\
CDLNet-S                & 652k & \underline{40.48} & 34.03      &     31.37  &     28.15 \\
BT CDLNet-S             & 669k & \underline{40.48} & \bf{34.07} & \bf{31.40} & \bf 28.19 \\\hline
\end{tabular}
}
\label{table:colorsingle_BT}
\end{table}

\begin{figure} 
    \centering
    \subfloat[PSNR plot $\sigma^{\mathrm{train}}={[01,20]}$ \label{fig:ColorBlindPlot0120BT}]{%
        \includegraphics[width=0.49\linewidth]{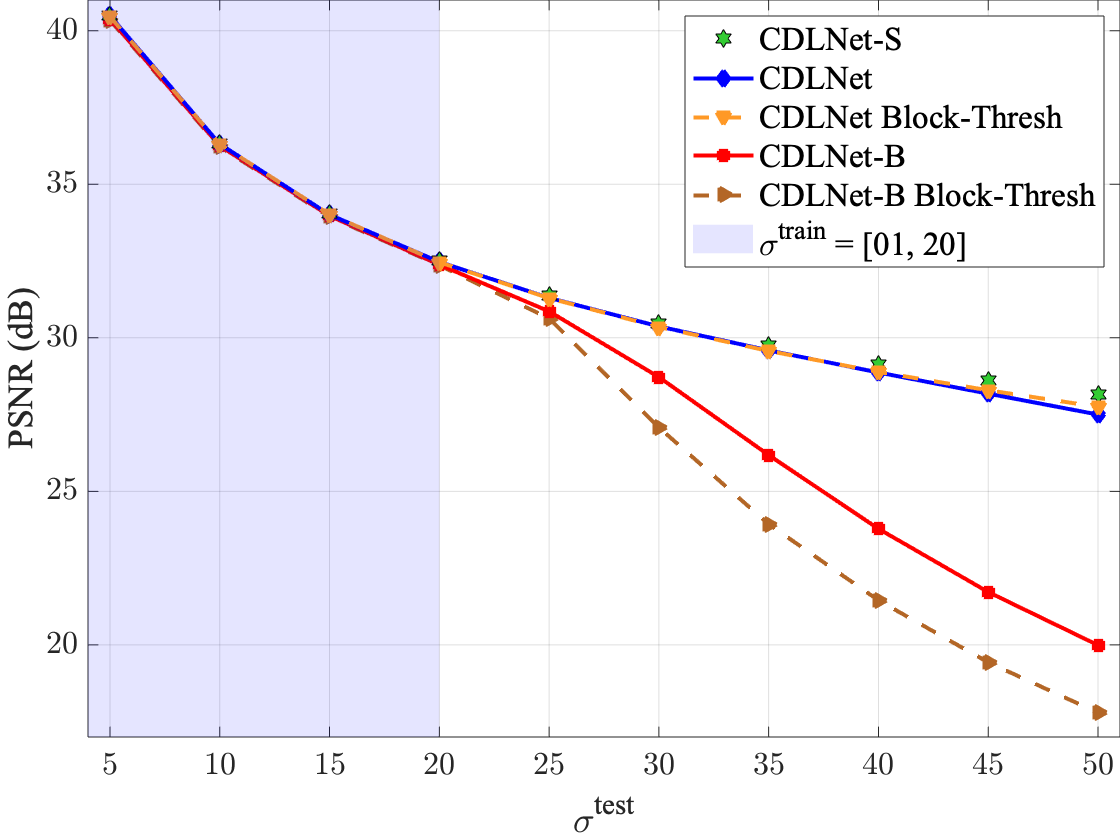}}
  \subfloat[PSNR plot $\sigma^{\mathrm{train}}={[20,30]}$ \label{fig:ColorBlindPlot2030BT}]{%
        \includegraphics[width=0.49\linewidth]{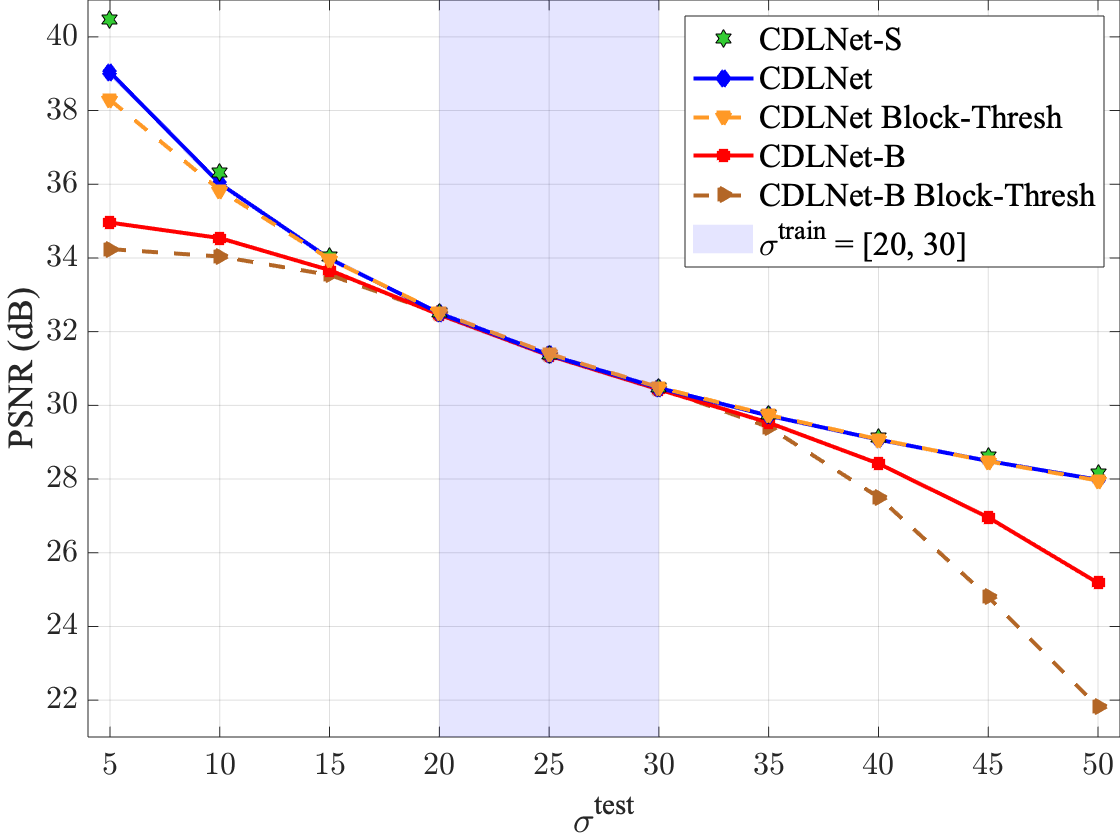}}
  \caption{(a,b) Performance of different color denoising networks trained on $\sigma^{\mathrm{train}}$ and tested on different $\sigma^{\mathrm{test}}$. Average PSNR calculated over CBSD68 \cite{bsd}.}
  \label{fig:ImgColorBT} 
\end{figure}

Figure \ref{fig:ImgColorBT} shows the generalization characteristics of the block-thresholding CDLNet (BT CDLNet) compared against its soft-thresholding counter-part (CDLNet) and associated networks without the proposed adaptive thresholding scheme (-B). Near-perfect generalization above and below the training noise-level range is also observed in the block-thresholding models. This suggests that the success of the proposed adaptive thresholding scheme is not unique to the specific architecture used, but rather a result of the proper interpretation of parameters from the convolutional dictionary learning derivation.

Lastly, Figure \ref{fig:BTDictionary} shows a comparison of the dictionaries from single noise-level and noise-level range trained color (a,d), block-thresholding (b,e), and grayscale (c,f) denoising networks. Similar to the color and grayscale models, the BT CDLNet models have qualitatively similar dictionaries whether trained on a single noise-level or noise-level range. Furthermore, we observe that the BT CDLNet models possess filters similar to those of both the grayscale models' filters and the grayscale filters of the color models.

\textbf{BT CDLNet vs. CDLNet}:
Ultimately, BT CDLNet yields a similar performance to CDLNet when scaled to similar parameter counts for the color denoising task. However when parameters are matched, BT CDLNet's forward pass requires 3 times the memory of CDLNet's, rendering training and inference much slower. This makes CDLNet clearly advantageous over BT CDLNet. 

We also trained a BT CDLNet on the JDD task and observed a significant performance decrease when compared to the presented JDD CDLNet. This is likely due to a modeling of color relationships being more necessary due to the RGGB Bayer mask in the JDD problem, which BT CDLNet cannot do. For this reason, and the aforementioned longer processing times, we omitted BT CDLNet from the main text in favor of CDLNet with a color-filter dictionary.

\subsection{Extended Data Figures and Tables}

Here, we present the values plotted in Figures \ref{fig:ImgGray} for gray-scale image denoising. The shaded areas on the tables show the training noise range. Similarly, we show values plotted in Figures \ref{fig:ImgColor} for color denoising and \ref{fig:ImgDMSC} for joint denoising and demosaicing task.

\begin{table}[ht]
\centering
\caption{Values in Fig. \ref{fig:GrayBlindPlot0120}. (Gray,  $\sigma^{\mathrm{train}}=[01,20],~ \sigma^{\mathrm{test}}\in[5,50]$)}
\resizebox{\linewidth}{!}{%
\begin{tabular}{ccccccccccc} \hline
\multirow{2}{*}{Model} & \multicolumn{10}{c}{Noise-level ($\sigma^{\mathrm{test}}$)} \\
 & 5 & 10 & 15 & 20 & 25 & 30 & 35 & 40 & 45 & 50 \\ \hline
CDLNet-S & 37.95 & 33.89 & 31.74 & 30.31 & 29.26 & 28.45 & 27.78 & 27.22 & 26.75 & 26.35 \\
CDLNet   & \cellcolor[rgb]{0.9,0.9,1}37.95 & \cellcolor[rgb]{0.9,0.9,1}33.88 & \cellcolor[rgb]{0.9,0.9,1}31.73 & \cellcolor[rgb]{0.9,0.9,1}30.29 & 29.23 & 28.38 & 27.69 & 27.09 & 26.57 & 26.09 \\
CDLNet-B &         \cellcolor[rgb]{0.9,0.9,1}37.73& \cellcolor[rgb]{0.9,0.9,1}33.74& \cellcolor[rgb]{0.9,0.9,1}31.59& \cellcolor[rgb]{0.9,0.9,1}30.13& 28.07& 24.37& 21.22& 18.97& 17.34& 16.09 \\
DnCNN-B$^{\ast}$ & \cellcolor[rgb]{0.9,0.9,1}37.81& \cellcolor[rgb]{0.9,0.9,1}33.78& \cellcolor[rgb]{0.9,0.9,1}31.61& \cellcolor[rgb]{0.9,0.9,1}30.21& 26.88& 23.30& 20.98& 19.29& 17.91& 16.76 \\
FFDNet-B$^{\ast}$& \cellcolor[rgb]{0.9,0.9,1}37.34& \cellcolor[rgb]{0.9,0.9,1}33.39& \cellcolor[rgb]{0.9,0.9,1}31.25& \cellcolor[rgb]{0.9,0.9,1}29.68& 28.09& 26.51& 25.93& 25.35& 24.90& 24.57 \\ \hline
\end{tabular}
}
\label{table:GrayBlindPlot0120Table}
\end{table}

\begin{table}[ht]
\centering
\caption{Values in Fig. \ref{fig:GrayBlindPlot2030}. (Gray,  $\sigma^{\mathrm{train}}=[20,30],~ \sigma^{\mathrm{test}}\in[5,50]$)}
\resizebox{\linewidth}{!}{%
\begin{tabular}{ccccccccccc} \hline
\multirow{2}{*}{Model} & \multicolumn{10}{c}{Noise-level ($\sigma^{\mathrm{test}}$)} \\
 & 5 & 10 & 15 & 20 & 25 & 30 & 35 & 40 & 45 & 50 \\ \hline
CDLNet-S &         37.95& 33.89& 31.74& 30.31& 29.26& 28.45& 27.78& 27.22& 26.75& 26.35 \\
CDLNet   &         36.74& 33.62& 31.67& \cellcolor[rgb]{0.9,0.9,1}30.30& \cellcolor[rgb]{0.9,0.9,1}29.25& \cellcolor[rgb]{0.9,0.9,1}28.44& 27.77& 27.19& 26.67& 26.22 \\
CDLNet-B &         31.14& 31.08& 30.85& \cellcolor[rgb]{0.9,0.9,1}30.16& \cellcolor[rgb]{0.9,0.9,1}29.16& \cellcolor[rgb]{0.9,0.9,1}28.33& 26.38& 22.65& 19.75& 17.74 \\
DnCNN-B$^{\ast}$ & 31.73& 31.50& 31.05& \cellcolor[rgb]{0.9,0.9,1}30.19& \cellcolor[rgb]{0.9,0.9,1}29.15& \cellcolor[rgb]{0.9,0.9,1}28.33& 25.81& 22.36& 19.97& 18.30 \\
FFDNet-B$^{\ast}$& 33.23& 32.43& 31.25& \cellcolor[rgb]{0.9,0.9,1}30.15& \cellcolor[rgb]{0.9,0.9,1}29.03& \cellcolor[rgb]{0.9,0.9,1}28.21& 26.45& 25.27& 24.31& 23.73 \\ \hline
\end{tabular}
}
\label{table:GrayBlindPlot2030Table}
\end{table}

\begin{table}[ht]
\centering
\caption{Values in Fig. \ref{fig:GrayBlindPlot0120Small}. (Gray,  $\sigma^{\mathrm{train}}=[01,20],~ \sigma^{\mathrm{test}}\in[5,50]$)}
\resizebox{\linewidth}{!}{%
\begin{tabular}{ccccccccccc} \hline
\multirow{2}{*}{Model} & \multicolumn{10}{c}{Noise-level ($\sigma^{\mathrm{test}}$)} \\
 & 5 & 10 & 15 & 20 & 25 & 30 & 35 & 40 & 45 & 50 \\ \hline
small-CDLNet-S &  37.90& 33.77& 31.60& 30.15& 29.11& 28.28& 27.63& 27.07& 26.59& 26.19\\
small-CDLNet   &  \cellcolor[rgb]{0.9,0.9,1} 37.87& \cellcolor[rgb]{0.9,0.9,1} 33.77& \cellcolor[rgb]{0.9,0.9,1} 31.58& \cellcolor[rgb]{0.9,0.9,1} 30.13& 29.06& 28.21& 27.50& 26.93& 26.41& 25.97\\
small-CDLNet-B &  \cellcolor[rgb]{0.9,0.9,1} 37.57& \cellcolor[rgb]{0.9,0.9,1} 33.62& \cellcolor[rgb]{0.9,0.9,1} 31.45& \cellcolor[rgb]{0.9,0.9,1} 29.96& 27.65& 23.89& 20.89& 18.85& 17.37& 16.19\\
CSCNet-B$^{\ast}$&  \cellcolor[rgb]{0.9,0.9,1} 37.55& \cellcolor[rgb]{0.9,0.9,1} 33.57& \cellcolor[rgb]{0.9,0.9,1} 31.40& \cellcolor[rgb]{0.9,0.9,1} 29.88& 27.91& 25.03& 22.41& 20.33& 18.64& 17.23\\
\hline
\end{tabular}
}
\label{table:GrayBlindPlot0120SmallTable}
\end{table}

\begin{table}[ht]
\centering
\caption{Values in Fig. \ref{fig:GrayBlindPlot2030Small}. (Gray,  $\sigma^{\mathrm{train}}=[20,30],~ \sigma^{\mathrm{test}}\in[5,50]$)}
\resizebox{\linewidth}{!}{%
\begin{tabular}{ccccccccccc} \hline
\multirow{2}{*}{Model} & \multicolumn{10}{c}{Noise-level ($\sigma^{\mathrm{test}}$)} \\
 & 5 & 10 & 15 & 20 & 25 & 30 & 35 & 40 & 45 & 50 \\ \hline
small-CDLNet-S &  37.90& 33.77& 31.60& 30.15& 29.11& 28.28& 27.63& 27.07& 26.59& 26.19\\
small-CDLNet   &  36.59& 33.52& 31.53&  \cellcolor[rgb]{0.9,0.9,1} 30.14&  \cellcolor[rgb]{0.9,0.9,1} 29.09&  \cellcolor[rgb]{0.9,0.9,1} 28.28& 27.60& 27.02& 26.51& 26.08\\
small-CDLNet-B &  30.65& 30.65& 30.56&  \cellcolor[rgb]{0.9,0.9,1} 30.05&  \cellcolor[rgb]{0.9,0.9,1} 29.04&  \cellcolor[rgb]{0.9,0.9,1} 28.18& 26.75& 23.98& 21.12& 18.90\\
CSCNet-B$^{\ast}$&  31.33& 31.16& 30.81&  \cellcolor[rgb]{0.9,0.9,1} 30.05&  \cellcolor[rgb]{0.9,0.9,1} 29.04&  \cellcolor[rgb]{0.9,0.9,1} 28.19& 26.96& 24.43& 21.62& 19.36\\
\hline
\end{tabular}
}
\label{table:GrayBlindPlot2030SmallTable}
\end{table}

\begin{table}[ht]
\centering
\caption{Values in Fig. \ref{fig:ColorBlindPlot0120}. (Color,  $\sigma^{\mathrm{train}}=[01,20],~ \sigma^{\mathrm{test}}\in[5,50]$)}
\resizebox{\linewidth}{!}{%
\begin{tabular}{ccccccccccc} \hline
\multirow{2}{*}{Model} & \multicolumn{10}{c}{Noise-level ($\sigma^{\mathrm{test}}$)} \\
 & 5 & 10 & 15 & 20 & 25 & 30 & 35 & 40 & 45 & 50 \\ \hline
CDLNet-S &          40.48& 36.33& 34.04& 32.51& 31.37& 30.48& 29.75& 29.13& 28.61& 28.15\\
CDLNet   &          \cellcolor[rgb]{0.9,0.9,1}40.49& \cellcolor[rgb]{0.9,0.9,1}36.31& \cellcolor[rgb]{0.9,0.9,1}34.01& \cellcolor[rgb]{0.9,0.9,1}32.47& 31.30& 30.37& 29.59& 28.87& 28.18& 27.50\\
CDLNet-B &          \cellcolor[rgb]{0.9,0.9,1}40.38& \cellcolor[rgb]{0.9,0.9,1}36.25& \cellcolor[rgb]{0.9,0.9,1}33.97& \cellcolor[rgb]{0.9,0.9,1}32.38& 30.86& 28.72& 26.18& 23.78& 21.72& 19.99\\
DnCNN-B$^{\ast}$ &  \cellcolor[rgb]{0.9,0.9,1}40.50& \cellcolor[rgb]{0.9,0.9,1}36.30& \cellcolor[rgb]{0.9,0.9,1}34.00& \cellcolor[rgb]{0.9,0.9,1}32.46& 31.12& 29.20& 27.09& 24.59& 22.51& 20.32\\
FFDNet-B$^{\ast}$&  \cellcolor[rgb]{0.9,0.9,1}40.03& \cellcolor[rgb]{0.9,0.9,1}35.92& \cellcolor[rgb]{0.9,0.9,1}33.61& \cellcolor[rgb]{0.9,0.9,1}32.42& 31.22& 29.87& 28.37& 27.19& 26.46& 25.73\\ \hline
\end{tabular}
}
\label{table:ColorBlindPlot0120Table}
\end{table}

\begin{table}[ht]
\centering
\caption{Values in Fig. \ref{fig:ColorBlindPlot2030}. (Color,  $\sigma^{\mathrm{train}}=[20,30],~ \sigma^{\mathrm{test}}\in[5,50]$)}
\resizebox{\linewidth}{!}{%
\begin{tabular}{ccccccccccc} \hline
\multirow{2}{*}{Model} & \multicolumn{10}{c}{Noise-level ($\sigma^{\mathrm{test}}$)} \\
 & 5 & 10 & 15 & 20 & 25 & 30 & 35 & 40 & 45 & 50 \\ \hline
CDLNet-S &          40.48& 36.33& 34.04& 32.51& 31.37& 30.48& 29.75& 29.13& 28.61& 28.15\\
CDLNet   &          39.04& 36.02& 33.98& \cellcolor[rgb]{0.9,0.9,1}32.50& \cellcolor[rgb]{0.9,0.9,1}31.37& \cellcolor[rgb]{0.9,0.9,1}30.47& 29.72& 29.07& 28.49& 27.98\\
CDLNet-B &          34.96& 34.54& 33.67& \cellcolor[rgb]{0.9,0.9,1}32.46& \cellcolor[rgb]{0.9,0.9,1}31.35& \cellcolor[rgb]{0.9,0.9,1}30.43& 29.53& 28.42& 26.96& 25.18\\
DnCNN-B$^{\ast}$ &  37.92& 35.71& 33.94& \cellcolor[rgb]{0.9,0.9,1}32.54& \cellcolor[rgb]{0.9,0.9,1}31.41& \cellcolor[rgb]{0.9,0.9,1}30.51& 29.66& 28.41& 26.85& 25.33\\
FFDNet-B$^{\ast}$&  38.42& 35.96& 33.94& \cellcolor[rgb]{0.9,0.9,1}32.54& \cellcolor[rgb]{0.9,0.9,1}31.39& \cellcolor[rgb]{0.9,0.9,1}30.49& 29.56& 28.76& 27.82& 26.81\\ \hline
\end{tabular}
}
\label{table:ColorBlindPlot2030Table}
\end{table}

\begin{table}[ht]
\centering
\caption{Values in Fig. \ref{fig:dmsc_generalize}. (JDD,  $\sigma^{\mathrm{train}}=[01,20],~ \sigma^{\mathrm{test}}\in[5,50]$)}
\resizebox{\linewidth}{!}{%
\begin{tabular}{ccccccccccc} \hline
\multirow{2}{*}{Model} & \multicolumn{10}{c}{Noise-level ($\sigma^{\mathrm{test}}$)} \\
 & 5 & 10 & 15 & 20 & 25 & 30 & 35 & 40 & 45 & 50 \\ \hline
JDD CDLNet-S & 36.36& 32.96& 31.00& 29.66& 28.66& 27.88& 27.23& 26.71& 26.24& 25.85 \\
JDD CDLNet   & \cellcolor[rgb]{0.9,0.9,1}36.29& \cellcolor[rgb]{0.9,0.9,1}32.95& \cellcolor[rgb]{0.9,0.9,1}30.98& \cellcolor[rgb]{0.9,0.9,1}29.62& 28.59& 27.77& 27.09& 26.51& 26.01& 25.54 \\
JDD CDLNet-B & \cellcolor[rgb]{0.9,0.9,1}36.08& \cellcolor[rgb]{0.9,0.9,1}32.82& \cellcolor[rgb]{0.9,0.9,1}30.86& \cellcolor[rgb]{0.9,0.9,1}29.44& 27.40& 23.84& 20.71& 18.54& 17.02& 15.86 \\
\hline
\end{tabular}
}
\label{table:dmsc_generalizeTable}
\end{table}

\subsection{Depth vs. width trade-off}
Here we examine the effect of architecture choices of number of channels (M), number of unrollings (K), and filter-size (P). Table \ref{tab:depth_width} suggests that favoring depth (unrollings) vs width (subbands and filter-size) will give improvement for a given parameter/computational budget (see second and third row of Table \ref{tab:depth_width}). Overall, the results show that the architecture is somewhat robust to hyperparameter tuning with the major indicator of performance being the learned parameter count. 

\begin{table}[htpp]
\caption{Depth vs. Width Trade-off}
\centering
\resizebox{0.85\linewidth}{!}{%
\begin{tabular}{ccccccc} \hline
$K$ & $M$ & $\sqrt{P}$ & $s$ & params & Flops/N & PSNR \\ \hline
50 & 49 &  7 & 2 & 245k & 30k  & 29.18 \\
30 & 169 & 5 & 2 & 264k & 32k  & 29.20  \\
10 & 289 & 7 & 2 & 289k & 35k  & 29.17 \\
20 & 169 & 7 & 2 & 338k & 41k  & 29.22 \\
30 & 121 & 7 & 2 & 363k & 44k  & 29.23  \\
30 & 169 & 7 & 2 & 507k & 62k  & 29.26 \\
30 & 225 & 7 & 2 & 675k & 83k  & 29.28 \\
40 & 169 & 7 & 2 & 676k & 83k  & 29.28 \\
30 & 169 & 9 & 2 & 831k & 103k & 29.28 \\
50 & 169 & 7 & 2 & 845k & 104k & 29.29 \\ 
30 & 289 & 7 & 2 & 867k & 106k & 29.29 \\
40 & 225 & 7 & 2 & 900k & 110k & 29.30 \\
40 & 448 & 7 & 2 & 1.8M & 220k & 29.34 \\ \hline
\end{tabular}
}
\label{tab:depth_width}
\end{table}

\subsection{Thresholds, filters, and sparse codes across layers}
\begin{figure}[hb] 
    \centering
    \subfloat[CDLNet $\sigma^{\mathrm{train}}={[20,30]}$ \label{fig:ABD}]{%
        \includegraphics[width=0.99\linewidth]{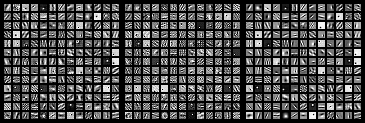}}
    \\\vspace*{-10pt}
     \subfloat[CDLNet-B $\sigma^{\mathrm{train}}={[20,30]}$\label{fig:ABDB}]{%
      \includegraphics[width=\linewidth]{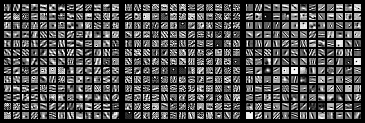}}
  \caption{(\textit{left}) Analysis ($\bm{A}^{(K)}$), (\textit{middle}) synthesis ($\bm{B}^{(K)}$), (\textit{right}) final synthesis ($\bm{D}$) dictionaries.}
  \label{fig:ImgDictLayers} 
\end{figure}

We visualized the learned dictionaries $\boldsymbol{A}^{(k)}$ and $\boldsymbol{B}^{(k)}$ for $k=0,\cdots,K$ in an animation available \href{https://nikopj.github.io/notes/cdlnet_supp/}{here}\footnote{\url{https://nikopj.github.io/notes/cdlnet_supp/}}. The last layer synthesis $\boldsymbol{A}^{(K)}$, analysis $\boldsymbol{B}^{(K)}$ and final dictionary $\bm{D}$ for CDLNet and CDLNet-B are shown in figures \ref{fig:ABD} and \ref{fig:ABDB}, respectively. We observed that as we progress through the layers of the network, the analysis and synthesis dictionaries look more Gabor-like and converge closer to the final dictionary. Interestingly, the very first few layers of the network also show Gabor-like structures, in contrast to the more ``noisy" filters in the intermediate layers. Further, we do not observe a significant difference between the dictionaries of CDLNet and CDLNet-B, suggesting that the generalization capability of CDLNet is solely a result of the noise-adaptive thresholds and not the learned intermediate representations. We have further investigated the Gabor-like nature of CDLNet's dictionaries in \cite{janjusevic2022gabor} by explicitly parameterizing filters as learnable 2D Gabor functions. The relationship of analysis and synthesis dictionaries over iterations is investigated by progressively untying Gabor parameters.



\begin{figure}[hb] 
    \centering
    \subfloat{%
        \includegraphics[width=0.99\linewidth]{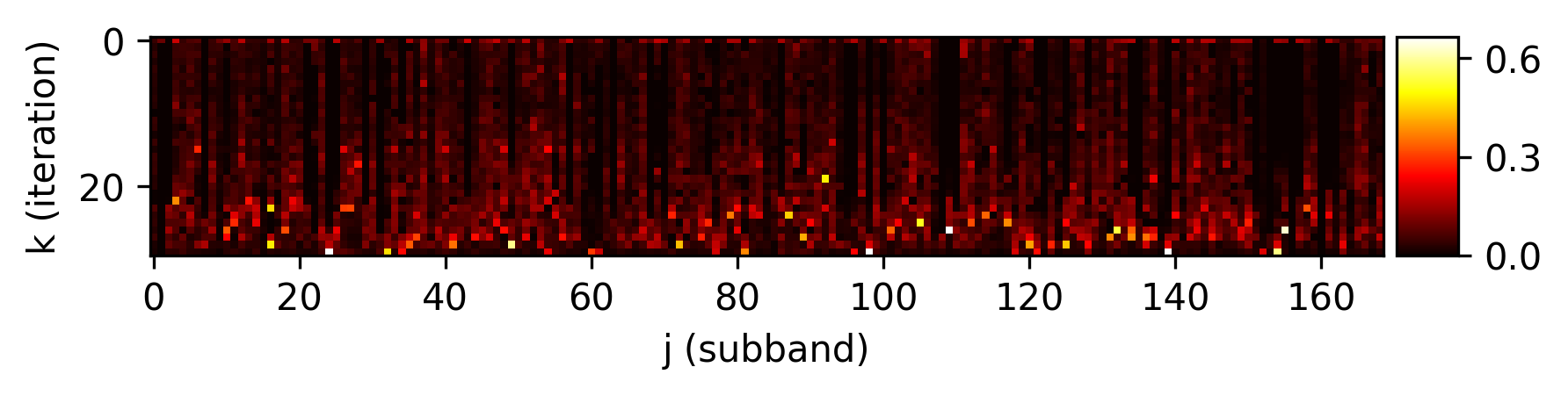}}
    \\\vspace*{-15pt}
     \subfloat{%
      \includegraphics[width=\linewidth]{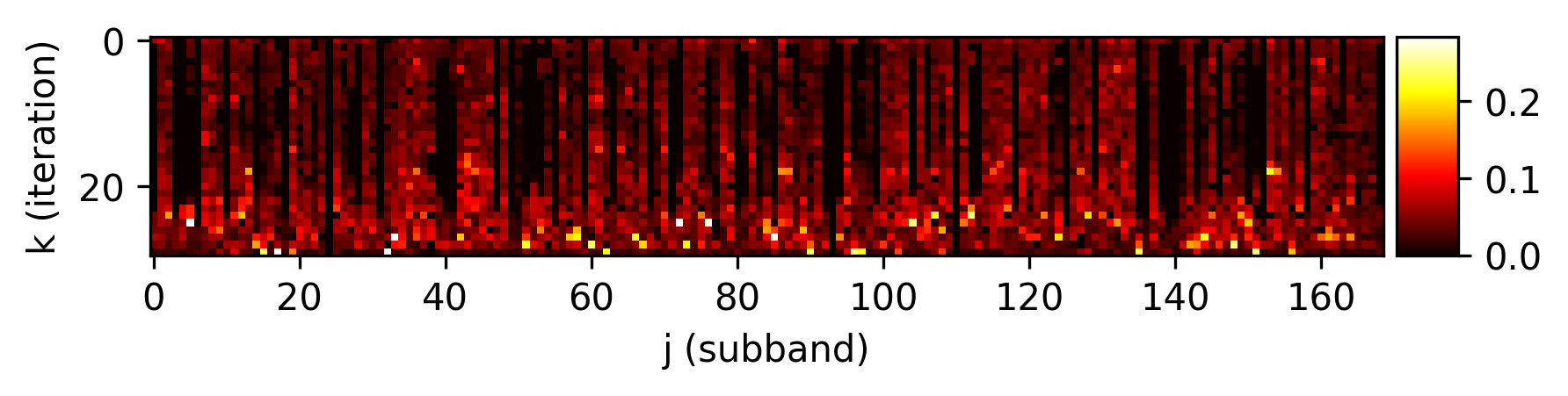}}
  \caption{Thresholds for CDLNet (top) and CDLNet-B (bottom) trained on $\sigma^{\mathrm{train}}=[20,30]$. CDLNet thresholds shown for $\sigma = 25$.  }
  \label{fig:Imgtau} 
\end{figure}

We visualized the learned thresholds for each layer $k=0,\dots ,K$ and subband $j=1,\dots,M$ as an image displayed in Figure \ref{fig:Imgtau}. For the adaptive model, we have an affine relationship with the input noise-level ($\tau^{(k)}=\tau^{(k)}_0+\tau^{(k)}_1\sigma$). For visualization purposes, we look at the thresholds for an input noise-level of $\sigma=25$. We also show the thresholds of an equivalent model trained without adaptive thresholds (CDLNet-B). For both adaptive and non-adaptive models, we see a general trend of thresholds increasing towards the final layers.

\begin{figure}[ht]
    \centering
    \includegraphics[width=0.7\linewidth]{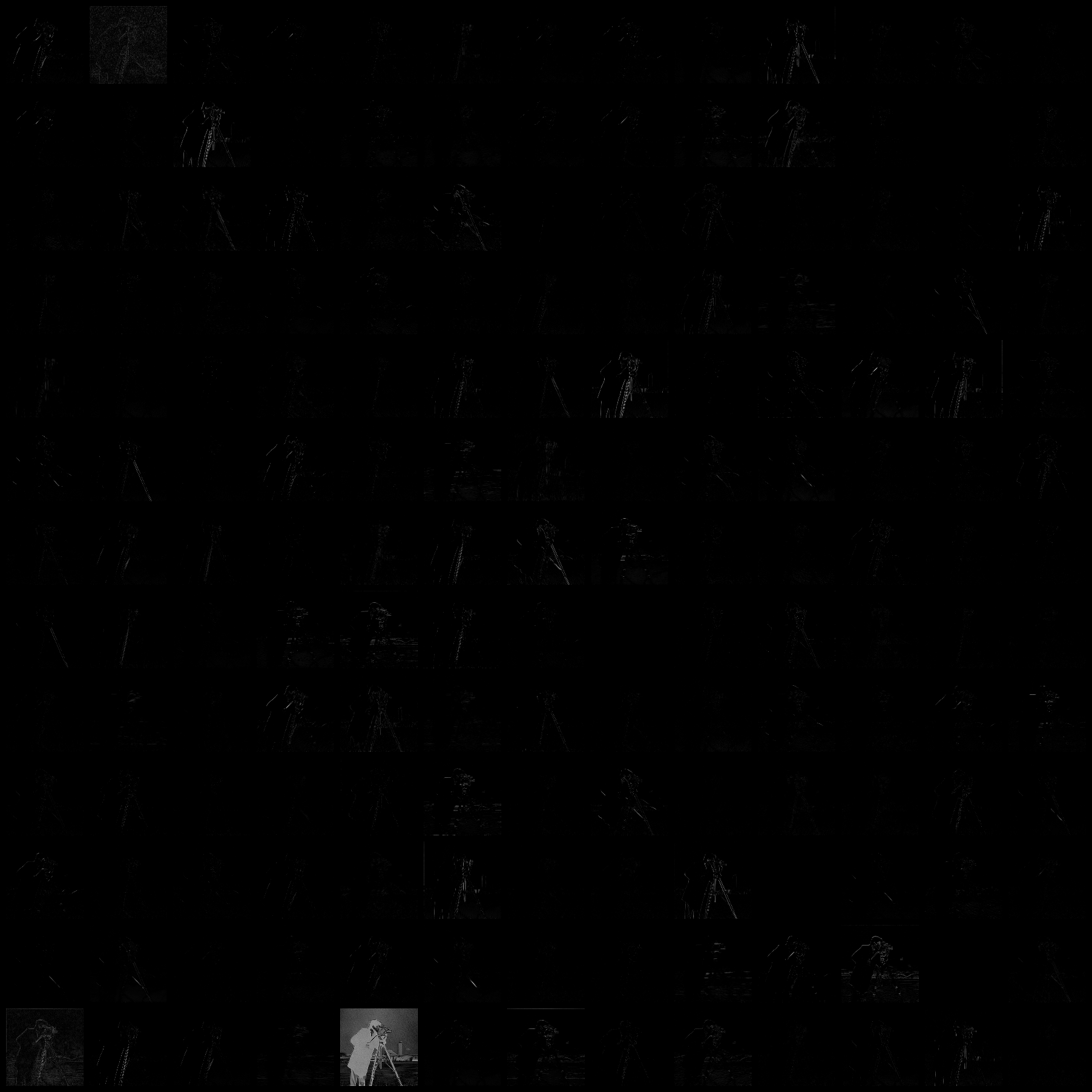}
    \caption{CDLNet trained over $\sigma^{\mathrm{train}}=[20,30]$. Final sparse codes ($\bm{z}^{(K)}$) for cameraman test image with noise-level $\sigma^\mathrm{test}=25$.}
    \label{fig:csc}
\end{figure}

We visualized the magnitude of the sparse codes for each layer (as an animation) for the cameraman test image, for an input noise-level of $\sigma=25$ \href{https://nikopj.github.io/notes/cdlnet_supp/}{here}. Figure \ref{fig:csc} shows the sparse codes for the last layer of the network. We observe that the representation becomes sparse in the final layers of the network, corresponding with the higher learned thresholds. Note that sparsity is not explicitly asked for during training (there is no sparsity penalty in the loss function), but rather it is encoded through the use of the shrinkage-thresholding non-linearity (derived from the basis-pursuit denoising formulation of the network).

\ifCLASSOPTIONcaptionsoff
  \newpage
\fi
\bibliographystyle{IEEEtran}

\bibliography{IEEEabrv,references}

\end{document}